\documentclass[12pt]{article}

\usepackage{hyperref}
\usepackage{graphicx}
\usepackage{amsthm,latexsym}
\usepackage{amsmath,amsfonts,amssymb,xspace,mathrsfs}
\usepackage{psfrag}
\usepackage{epsfig}
\usepackage{url}

\usepackage{hyperref}

\allowdisplaybreaks
\numberwithin{equation}{section}

\newcounter{tabl}
\newcommand{\be}{\begin{equation}}
\newcommand{\ee}{\end{equation}}
\def\ba{\begin{eqnarray}}
\def\ea{\end{eqnarray}}
\newcommand{\beq}{\begin{eqnarray}}
\newcommand{\eeq}{\end{eqnarray}}
\newcommand{\bea}[2]{\be\label{#2}\begin{array}{#1}}
\newcommand{\eea}{\end{array}\ee}

\def\tr{\,{\rm Tr}\, }
\def\det{\,{\rm det}\, }
\def\diag{{\rm diag}}

\def\tr{\,{\rm tr}\, }

\def\({\left(}
\def\){\right)}
\def\[{\left[}
\def\]{\right]}
\def\p{\partial}
\newcommand{\de}{\mathrm{d}}
\newcommand{\I}{\mathrm{i}}

\def\11{1\!\! 1}

\def\hf{\frac{1}{2}}

\def\eps{\varepsilon}

   \def\CC {{\cal C}}
   \def\CD {{\cal D}}

   \def\CG {{\cal G}}
   \def\CH {{\cal H}}
   \def\CI {{\cal I}}

   \def\CM {{\cal M}}

   \def\CQ {{\cal Q}}
   
   \def\CS {{\cal S}}

   \def\CV {{\cal V}}

   \def\CZ {{\cal Z}}

\newcommand{\unit}{\hbox{$1\hskip -2.2pt\vrule depth 0pt height 1.6ex width 0.7pt
                  \vrule depth 0pt height 0.3pt width 0.12em$}}

\newcommand{\Cmat}{{\mathbb C}}

\newcommand{\Rmat}{{\mathbb R}}

\newcommand{\gmat}{\mathfrak{g}}

\newcommand{\im}{\gamma}

\newcommand{\tP}{\lefteqn{\smash{\mathop{\vphantom{<}}\limits^{\;\sim}}}P}
\newcommand{\tG}{\lefteqn{\smash{\mathop{\vphantom{<}}\limits^{\;\sim}}}\CG}
\newcommand{\tp}{\lefteqn{\smash{\mathop{\vphantom{\scriptstyle{c}}}\limits^{\sim}}}p}

\newcommand{\pt}{\lefteqn{\smash{\mathop{\vphantom{\biggl(}}\limits_{\sim}
\atop \ }}\,p}

\newcommand{\tPb}[2]{{\tP_{(\im)#2}^{#1}}}

\newcommand{\tPp}[2]{{\tP^{(+)#1}_{#2}}}
\newcommand{\tPm}[2]{{\tP^{(-)#1}_{#2}}}
\newcommand{\tPpm}[2]{{\tP^{(\pm)#1}_{#2}}}

\newcommand{\Bp}[2]{{B^{(+)#2}_{0#1}}}
\newcommand{\Bm}[2]{{B^{(-)#2}_{0#1}}}

\newcommand{\gb}{{\rm g}}

\newcommand{\gl}{g}
\newcommand{\Db}{{\bf D}}

\def\Int{\CI}
\def\inter{i}

\def\Tr{\text{Tr}}

\newcommand{\SU}{\text{SU}}

\newcommand{\Sp}{\text{Spin}}
\newcommand{\su}{\mathfrak{su}}

\newcommand{\so}{\mathfrak{so}}

\def\xc{{\rm x}}
\def\yc{{\rm y}}
\def\ccon{\Upsilon}

\newcommand{\ClG}[6]{{{C\lefteqn{\vphantom{\overline{A}_-}}^{\smash{#1}}_{\smash{#4}}}^{\smash{#2}}_{\smash{#5}}}^{{#3}}_{{#6}}}

\def\xb{\mathrm{x}}
\def\eb{\mathbf{e}}
\def\gf{\mathrm{g}}

\def\piI{\pi_{(1)}}
\def\piII{\pi_{(2)}}

\begin{document}

\thispagestyle{empty}

\

\vspace{1.3cm}

\begin{center}
{\LARGE \bfseries
Degenerate Plebanski Sector
}
\\
\
\\
{\LARGE \bfseries and Spin Foam Quantization}

\vskip 1cm

{\Large Sergei Alexandrov}

\vskip 0.5cm

%

{\it Universit\'e Montpellier 2 \& CNRS, Laboratoire Charles Coulomb \\
UMR 5221, F-34095, Montpellier, France}
\end{center}

\vskip 0.6cm

\centerline{\bf Abstract}

\begin{quote}\small
We show that the degenerate sector of Spin(4) Plebanski formulation of four-dimensional gravity
is exactly solvable and describes covariantly embedded SU(2) BF theory.
This fact ensures that its spin foam quantization is given by the SU(2) Crane-Yetter model
and allows to test various approaches of imposing the simplicity constraints.
Our analysis strongly suggests that restricting representations and intertwiners in the state sum for $\Sp(4)$ BF theory
is not sufficient to get the correct vertex amplitude.
Instead, for a general theory of Plebanski type, we propose a quantization procedure which is
by construction equivalent to the canonical path integral quantization and, being applied to our model,
reproduces the SU(2) Crane-Yetter state sum. A characteristic feature of this procedure is
the use of secondary second class constraints on an equal footing with the primary simplicity constraints,
which leads to a new formula for the vertex amplitude.
\end{quote}


\newpage
\pagenumbering{arabic}

\tableofcontents

\section{Introduction}

Spin foam quantization appears as an attempt to construct
a well-defined path integral for quantum gravity, representing it as
a sum over two-dimensional cellular complexes colored by ceratin group theoretic data \cite{Baez:1997zt,Perez:2003vx}.
Whereas in three-dimensions it is the full-fledged approach \cite{Ponzano:1968dq,Turaev:1992hq},
which can be related to other quantization schemes \cite{Witten:1988hf,Reshetikhin:1991tc,Noui:2004iy}
based on a solid canonical analysis,
in four dimensions its status is much more controversial.
The only models which are affordable to the
direct spin foam quantization are the so-called BF theories \cite{Horowitz:1989ng}.
These are topological theories described by the following action
\be
S_{\rm BF}[\omega,B]=\int_{\CM}\de^4\xc\,  \Tr\(B\wedge F(\omega)\),
\label{BFact}
\ee
where $\omega$ is the connection one-form valued in the Lie algebra $\gmat$ of a certain group $G$,
$F(\omega)$ is its curvature two-form, $B$ is a two-form also valued
in $\gmat$, and $\Tr$ is evaluated using the Killing form on $\gmat$.
The spin foam quantization of the theory \eqref{BFact} is well-known and given by
the Crane-Yetter model \cite{Crane:1993if} with the structure group $G$
(for recent developments, see also \cite{Bonzom:2012mb}).

Besides these topological spin foam models of BF theories, there have been suggested several
models supposed to describe gravity. Among them the most prominent ones are the Barrett-Crane (BC) model
\cite{Barrett:1997gw,Barrett:1999qw} and the so-called new models, which are due to
Engle, Pereira, Rovelli, Livine (EPRL) \cite{Engle:2007wy} and
Freidel, Krasnov (FK) \cite{Freidel:2007py,Livine:2007ya}.
However, all of them have been derived using certain assumptions and simplifications.
Their common starting point is the so-called Plebanski theory,
which represents gravity as a constrained BF theory
with $G=\Sp(4)$ in the Riemannian or $G=SL(2,\Cmat)$ in the Lorentzian case.
Namely, it realizes the simple fact that, if the action \eqref{BFact} is supplemented by a constraint ensuring that
\be
B=\star(e\wedge e),
\label{Bee}
\ee
it reproduces the usual Hilbert-Palatini formulation.
The idea leading to the spin foam models mentioned above is that these constraints, called commonly simplicity,
can be incorporated at the quantum level. Thus, they are supposed to be imposed
on the spin foam representation of the quantum BF theory and should convert the trivial dynamics of a topological theory
into that of quantum gravity. This strategy, which can be summarized as ``first quantize, then constrain",
is now the usual approach to four-dimensional spin foam models,
and what distinguishes various models is only the way the simplicity constraints are incorporated.

However, although this strategy seems to be well motivated and leads to interesting results, it does not agree
with the rules of quantization of constraint systems. The simplicity constraints are known to be second class
and therefore affect the symplectic structure to be quantized, which can be evaluated using the Dirac bracket.
It has been argued that this and some other effects are not taken properly into account in the spin foam approach
based on the above strategy \cite{Alexandrov:2010pg,Alexandrov:2010un}.
In particular, relying on the consistency with the canonical quantization,
it has been suggested a certain modification of the vertex amplitude \cite{Alexandrov:2008da},
which is the most important quantity in spin foam models encoding their dynamics.
If one follows the usual strategy, the vertex coincides with the one of BF theory, but restricted
to a set of representations and intertwiners, assigned to the elements of the cellular decomposition,
satisfying the simplicity constraints. Equivalently, it can be represented as the boundary state
associated with a four-simplex evaluated on a flat connection, or as
\be
A_v=\(\prod_\tau\int_G \CD \gl_\tau\)
\Psi\[\gl_{u(f)}^{-1}\gl_{d(f)}^{\mathstrut}\],
\label{amplit_simplex-gen}
\ee
where $\Psi[g_f]$ is the boundary state depending on the group elements assigned to triangles of the four-simplex,
the product goes over its tetrahedra, $u(f)/d(f)$ denotes the upper/down tetrahedron sharing triangle $f$,
and the integration measure should be taken to be the usual Haar measure on the group, $\CD\gl=\de\gl$.
In \cite{Alexandrov:2008da} it has been argued that the correct vertex for the constrained theory
should be given by the same formula,
but with the measure which involves the delta-function of secondary second class constraints,
conjugate to the simplicity and ensuring that they are of second class.

Given this situation when there are several proposals for the spin foam quantization of quantum gravity,
it is desirable to have some simplified models which allow to test various features of these proposals.
Moreover, since most problems and ambiguities arising in four-dimensional spin foam models come from
the difficulties in imposing the simplicity constraints,
such a model should mimic the structure of Plebanski formulation. In other words, it should
be of the following form
\be
S_{\rm Th2}=S_{\rm Th1}+{\rm constraints},
\label{formact}
\ee
where the constraints convert the theory Th1 into the theory Th2. Finally, to be a good
test-ground, both theories, given by the actions $S_{\rm Th1}$ and $S_{\rm Th2}$,
should have known spin foam representations.
Then we can verify which of the methods to impose the constraints reduces the spin foam quantization of
${\rm Th1}$ to that of ${\rm Th2}$.
If a method does not allow to recover the known quantization of Th2, this strongly suggests
that it is not applicable also in the case of Plebanski theory.
On the other hand, if we find a quantization procedure which passes through our test,
one can hope that it will work for gravity as well.

In four dimensions the above requirements suggest that Th1 and Th2 should be of BF type \eqref{BFact}
because these are the only theories with a well-established spin foam quantization.
Then the constraints can be used to reduce a gauge group $G$ to its subgroup $H\subset G$.
Thus, one arrives at an essentially unique model suitable for all our purposes:
it should represent BF theory with the gauge group, for example, $\Sp(4)$ reduced by means
of some constraints to the $\SU(2)$ BF theory.
The remaining question is which constraints in \eqref{formact} can ensure such a reduction?

In fact, similar models have already appeared in the literature.
First, in \cite{Alexandrov:2008da} it was shown that the above reduction does take place if
one imposes {\it two} constraints restricting the $B$-field and the spin connection, respectively.
They were supposed to arise as primary and secondary second class constraints from a certain action.
Under this assumption, it was demonstrated that one recovers the known spin foam representation
of the $\SU(2)$ BF theory from the known spin foam quantization of the $\Sp(4)$ BF theory
only if one incorporates the constraints on the spin connection
into the definition of the vertex amplitude in the way described below Eq. \eqref{amplit_simplex-gen}.
However, the important drawback of this consideration was that the constraints
have been imposed by hand and not derived from a classical action of Plebanski type.
On the other hand, an analogous model, but based on a solid canonical analysis, was proposed recently in \cite{Geiller:2011kr}
(see also \cite{Alexandrov:2011km}).
It also represents the reduction of $\Sp(4)$ BF to $\SU(2)$ BF, but this time in three dimensions
where the resulting theory is nothing else but three-dimensional gravity with vanishing cosmological constant.
Its analysis has led to the same conclusion that the simplicity constraints should be supplemented
by the secondary second class constraints restricting the holonomies of the spin connection.
But the three-dimensional nature of this model raises the question whether it is actually able to capture
all features of the constraint imposition in four-dimensions.

Taking these issues into account, in this paper we return to the original construction of \cite{Alexandrov:2008da}
and provide a full-fledged model of the type \eqref{formact}, which we carefully analyze both at classical and quantum level.
Furthermore, relying on this analysis, we propose a quantization procedure to build the spin foam partition function
for a general theory of Plebanski type, which therefore should be applicable to the gravity case as well.

The organization of the paper is as follows.
In section \ref{sec_Pleb} we propose a classical action which represents the $\Sp(4)$ BF theory
reduced down to $\SU(2)$ BF, where the constraints of \cite{Alexandrov:2008da} appear as primary
and secondary constraints, respectively. Remarkably, this action represents a system very close
to the physical system we are interested in --- it describes the degenerate sector of Plebanski theory.
Thus, the latter is exactly solvable and can be seen as a covariant embedding of the well known topological theory.
We provide a thorough canonical analysis of this degenerate sector both with a partial gauge fixing and without it,
including also the Immirzi parameter which does not lead to any complications.

Then in section \ref{sec-quant} we consider the spin foam quantization of this model.
First, we apply the usual quantization strategy employed in the EPRL and FK approaches.
Since our classical action differs from Plebanski theory only by the presence of the degeneracy condition,
to get its spin foam quantization, it is sufficient to extract the degenerate sector of the new spin foam models.
If it is done, following the usual ideas, by restricting the boundary or kinematical data,
the result strongly disagrees with the known vertex amplitude for the four-dimensional $SU(2)$ BF theory.
On the other hand, if the degeneracy condition is represented as a constraint on the group elements
appearing in the integral formula \eqref{amplit_simplex-gen} for the vertex and is added to the integration measure,
one does get the right result. Since this constraint can be equally seen as a discretization of
the secondary second class constraints, this modification of the measure is in the full agreement with our proposal
for the vertex amplitude spelled above.

At the same time, the analysis of the constraint imposition in the framework of the new models
shows that the constraint on holonomies ensuring the correct vertex amplitude erases all information about
the solution of the original simplicity constraints of the EPRL and FK models.
Thus, the main ingredients of these models appear to be irrelevant for getting the right dynamics,
which calls for a reconsideration of these approaches.
As an alternative, we suggest another quantization procedure, which summarizes
the analysis of \cite{Alexandrov:2008da,Alexandrov:2011km} and is consistent with
the canonical approach by construction.
Being applied to the model under consideration, it gives precisely the right result:
the Crane-Yetter model with the $SU(2)$ structure group.
Moreover, in the course of evaluation of the partition function, we clarify
the role of different constraints in the spin foam quantization. In particular, we observe that
the vertex amplitude is completely determined by the secondary second class constraints
putting restrictions on holonomies of the spin connection, whereas the primary simplicity constraints affect
only the gluing of different vertex contributions and are not relevant for dynamics.
A discussion of these and other issues can be found in the concluding section.

Our conventions are explained in appendix \ref{ap-conven}.
We restrict ourselves to the Riemannian case to not bother the reader with signs
which otherwise would pop out here and there.
However, all the presented results are easily generalized to the Lorentzian case as well.
In appendix \ref{ap-canan} we present the details of the canonical analysis
of the degenerate Plebanski sector without a partial gauge fixing,
whereas the last appendix \ref{ap-constr} makes explicit
the constraints for constraints.

\section{Degenerate sector of Plebanski formulation}
\label{sec_Pleb}

\subsection{The action and constraints}
\label{subsec-action}

Let us start from the usual Plebanski action
\be
\label{plebanskiaction2}
S_\text{dPl}[\omega,B,\lambda]=\hf\int_\CM\de^4\xc\left[\eps^{\mu\nu\rho\sigma}\Tr(B_{\mu\nu}F_{\rho\sigma})
+\frac{1}{2}\,\lambda^{\mu\nu\rho\sigma}\Tr(\star B_{\mu\nu}B_{\rho\sigma})\right],
\ee
where the Lagrange multiplier field $\lambda$ is chosen to be a spacetime pseudo-tensor
satisfying the following symmetry properties: it is antisymmetric in the first and second pair
of indices, $\lambda^{\mu\nu\rho\sigma}=\lambda^{[\mu\nu][\rho\sigma]}$,
and is symmetric under their exchange, $[\mu\nu]\leftrightarrow [\rho\sigma]$.
Usually, one also adds the tracelessness condition
$\eps_{\mu\nu\rho\sigma}\lambda^{\mu\nu\rho\sigma}=0$ \cite{Plebanski:1977zz,DePietri:1998mb} which we however omit.
The absence of this condition on $\lambda$ is the feature which distinguishes our model
from the usual Plebanski formulation of general relativity and is responsible for its solvability.
As a result, the variation with respect to the Lagrange multiplier generates
the following constraints
\be
\label{simplicity constraint}
\Phi_{\mu\nu\rho\sigma}=\hf\,\eps_{IJKL}B^{IJ}_{\mu\nu}B^{KL}_{\rho\sigma}=0.
\ee
They represent the usual 20 simplicity constraints supplemented by an additional condition
which forces the $B$-field to belong to the degenerate sector, i.e. to give a vanishing four-dimensional
volume $\CV=\frac{1}{12}\,\eps^{\mu\nu\rho\sigma}\Phi_{\mu\nu\rho\sigma}$.
However, in contrast to the usual case, not all of these 21 constraints are independent.
It turns out that there are 6 constraints for constraints, which we display
explicitly in appendix \ref{ap-constr}.
They are responsible for the well known fact \cite{Dittrich:2008ar,Conrady:2008mk}
that the phase space of degenerate configurations is larger than its non-degenerate version.
As a result, one remains only with 15 independent constraints
which thus reduce the number of independent components of the $B$-field
from 36 to 21.

It is easy to see that
\begin{subequations}
\beq
\mbox{``deg-gravitational" :}\quad\quad B_{\mu\nu}^{IJ}&=&\hf\,{\eps^{IJ}}_{KL}x^K b_{\mu\nu}^L,
\label{gravsector}
\\
\mbox{``deg-topological" :}\quad\quad B_{\mu\nu}^{IJ}&=&x^{[I} b_{\mu\nu}^{J]},
\label{tpsector}
\eeq
\label{twosectors}
\end{subequations}
where the vector $x^I$ is supposed to be normalized as $x^I x_I=1$,
are solutions to the simplicity constrains \eqref{simplicity constraint}. They represent two disjoint sectors
which we call ``degenerate gravitational" and ``degenerate topological",
in analogy with the usual case.\footnote{There might be also a ``twice degenerate" sector because
the proof in appendix \ref{ap-constr} that there are only 6 constraints for constraints
relies on the assumption that a certain matrix constructed from the $B$-field is invertible.
But we do not consider here this possibility and assume that the field $b_{\mu\nu}^I$ is generic so that
it ensures certain non-degeneracy conditions appearing in the course of our analysis.}
The reason for this will be clear from what follows.
Both solutions \eqref{twosectors} contain 21 independent degrees of freedom, i.e. the same as the solution space
of the simplicity constraints: 3 degrees of freedom are contained in $x^I$ and 18 are described by $b_{\mu\nu}^I$
because the latter field can be chosen to satisfy the linear constraints
\be
x_I b_{\mu\nu}^I=0.
\label{linsimpl}
\ee
They fix uniquely the ambiguity in $b_{\mu\nu}^I$ and will be always assumed to hold
in the following analysis.

If one fixes the vector $x^I$, thereby reducing the gauge symmetry from $\Sp(4)$ to the subgroup $\SU(2)_x$
which preserves this vector, the two sectors of solutions \eqref{twosectors} can be equivalently
characterized by {\it linear simplicity constraints}
\begin{subequations}
\beq
\mbox{``deg-gravitational" :}&\quad& \Phi^{({\rm gr})I}_{\mu\nu}=x_J B_{\mu\nu}^{IJ}=0,
\label{gravsectorlin}
\\
\mbox{``deg-topological" :}&\quad& \Phi^{({\rm top})I}_{\mu\nu}={\eps^{IJ}}_{KL}x_J B_{\mu\nu}^{KL}=0,
\label{tpsectorlin}
\eeq
\label{twosectorslin}
\end{subequations}
again in direct analogy with the usual non-degenerate case
\cite{Alexandrov:2007pq,Engle:2007qf,Freidel:2007py,Gielen:2010cu}.\footnote{Note however that
in the non-degenerate case the roles of $\Phi^{({\rm gr})}$ and $\Phi^{({\rm top})}$ are exchanged,
i.e. the former corresponds to the topological sector and the latter to the gravitational.\label{foot-twoconstr}}
However, the difference is that here the linear constraints \eqref{twosectorslin} exhaust {\it all} simplicity constraints,
whereas in the non-degenerate case they should be supplemented by the volume constraint.
Indeed, since both constraints \eqref{twosectorslin} satisfy
$x_I  \Phi^{({\rm gr})I}_{\mu\nu}=x_I  \Phi^{({\rm top})I}_{\mu\nu}=0$, they
give rise to 18 independent equations. This reduces the number of independent components
of the $B$-field to 18, which coincides with the number of independent degrees of freedom
described by $b_{\mu\nu}^I$.

To understand the meaning of the two sectors, let us assume for simplicity that $x^I=const$.
Then one can easily extract an equation without derivatives
from the equation of motion obtained by variation of \eqref{plebanskiaction2}
with respect to the spin connection
\be
\eps^{\mu\nu\rho\sigma}D_{\nu}B_{\rho\sigma}^{IJ}=0,
\label{eqtors}
\ee
where $D_\mu$ is the covariant derivative defined by $\omega_\mu$.
Indeed, let us contract the equation \eqref{eqtors} with $x_J$ in the deg-gravitational sector and with
${\eps^{KL}}_{IJ}x_L$ in the deg-topological sector. Taking into account the form of the $B$-field \eqref{twosectors},
in both cases one finds
\be
\eps^{\mu\nu\rho\sigma}{\eps^{IJ}}_{KL}x_J b_{\rho\sigma}^K(\omega_{\nu}^{LM}x_M) =0.
\ee
Under the assumption of invertibility of the matrix multiplying the spin connection, or more precisely $x_J\omega_{\mu}^{IJ}$,
this equation requires that
\be
x_J\omega_{\mu}^{IJ}=0 \quad \Longrightarrow\quad  x_J F_{\mu\nu}^{IJ}=0.
\label{constr-om}
\ee
This result indicates that only the part of the connection describing the $\SU(2)_x$ subgroup
survives, whereas the orthogonal part vanishes.
Now we can plug the solution of the simplicity constraints \eqref{twosectors} into the original action \eqref{plebanskiaction2}.
It is immediate to see that, due to \eqref{constr-om}, in the deg-topological sector the resulting action
identically vanishes, whereas in the deg-gravitational sector it becomes
\be
S_\text{gr}[\omega,b]=\frac14\int_\CM\de^4\xc\,\eps^{\mu\nu\rho\sigma}\eps_{IJKL}x^I b_{\mu\nu}^J F_{\rho\sigma}^{KL}.
\label{resactsol}
\ee
This is nothing else but the action of the four-dimensional $\SU(2)$ BF theory covariantly embedded into $\Sp(4)$.
Fixing the time gauge $x^I=\delta^I_0$, one recovers its usual form written in terms of the $\su(2)$-valued
2-form $b_{\mu\nu}^i$ and the $\su(2)$-connection $\omega_\mu^{ij}$. The other components of the original fields vanish
due to \eqref{linsimpl} and \eqref{constr-om}.

Since in the deg-topological sector we do not obtain any meaningful theory, we will be mainly concentrated
on the deg-gravitational sector. Remarkably, it is given by a well known topological theory
so that we know its exact classical as well as quantum description.
For our purposes it will be however important to understand also the canonical structure of the original
$\Sp(4)$-invariant theory \eqref{plebanskiaction2} which we present in the next subsection.
Then in subsection \ref{subsec-reduct} we will see how the reduction to
the $\SU(2)$ BF theory described here is established in the Hamiltonian formalism.
The reader who is not interested in details of this canonical analysis can proceed directly to section \ref{subsec-sum}.

\subsection{Canonical analysis}
\label{subsec-canan}

Here we present the Hamiltonian formulation of the degenerate gravitational sector of Plebanski theory
in a partially fixed gauge. Namely, we fix the normal field $x^I(\xc)$ to be a given function on
spacetime.
The choice of the normal allows to replace the quadratic simplicity constraints \eqref{simplicity constraint}
by their linearized version \eqref{twosectorslin}, which can be done directly in the action and gives the possibility
to restrict to the particular sector   we are interested in from the very beginning.
An analogous formulation of the gravitational sector of Plebanski theory has been considered in \cite{Gielen:2010cu}
(see also \cite{Gielen:2011mk}).
However, the important difference of our approach is that the normal $x^I$ is considered as a fixed
{\it non-dynamical variable}.
This is motivated by the following application of these results to the spin foam quantization of our model.
This quantization is implemented via a path integral where
the gauge freedom generated by boost transformations should be fixed by a gauge choice.
The most convenient way to do this is precisely to fix the normal $x^I$.
This gauge fixing is analogous to what is done in the standard loop quantum gravity where one imposes the so-called
time gauge corresponding to a particular choice of $x^I=\delta^I_0$.
Here we could also restrict ourselves to this simple gauge, in which case
the following derivation considerably simplifies. We however prefer to keep $x^I$
an arbitrary function to show that in this general case one obtains nice covariant structures.

We emphasize that the results presented here can also be derived going through the complete canonical analysis
of the original action \eqref{plebanskiaction2} carried out without imposing any gauge fixing.
We provide such analysis in appendix \ref{ap-canan} where the canonical structure of both
solution sectors \eqref{twosectors} is elucidated.
They possess a very intricate constraint structure which however reduces to the one
of this subsection upon restricting to the deg-gravitational sector, fixing $x^I$, and solving
auxiliary constraints.

Thus, our starting point is the following action
\be
\label{actiongauge}
S_\text{deg-gr}[\omega,B,\lambda;x]=\hf\int_\CM\de^4\xc\left[\eps^{\mu\nu\rho\sigma}B_{\mu\nu}^{IJ}F_{\rho\sigma}^{IJ}
+4\lambda^{\mu\nu}_I x_J B_{\mu\nu}^{IJ}\right],
\ee
where $x^I$ is just a parameter and not a dynamical variable.
After the 3+1 decomposition of this action, one can recognize that the phase space is parametrized by
\be
\omega_{a}^{IJ}
\qquad
{\rm and}
\qquad
\tP^a_{IJ}=\eps^{abc}B_{bc}^{IJ}
\label{dynvar}
\ee
with the canonical commutation relations
\be
\{\omega_a^{IJ}(\xc), \tP^b_{KL}(\yc)\}=\delta_a^b \delta^{IJ}_{KL}\delta(\xc,\yc).
\label{Poisbr}
\ee
The variables $\omega_{0}^{IJ}$, $B_{0a}^{IJ}$ and $\lambda^{\mu\nu}_I$ appear without time derivatives
and therefore play the role of Lagrange multipliers. However, not all of them generate constraints.
Whereas the variation with respect to $\omega_{0}^{IJ}$ and $\lambda^{ab}_I$ does give rise to
the primary constraints
\begin{subequations}
\beq
\CG_{IJ}&=& D_a \tP^a_{IJ}\approx 0,
\\
\Phi^{a}_{I}&=&x^J \tP^a_{IJ}\approx 0,
\label{Phi-x}
\eeq
\label{primarygauge}
\end{subequations}
the variation with respect to $\lambda^{0a}_I$ leads to a condition on the Lagrange multipliers,
\be
x_J B_{0a}^{IJ}=0.
\label{B0-x}
\ee
At the same time, the variation with respect to $B_{0a}^{IJ}$ gives the following equation
\be
\eps^{abc} F_{bc}^{IJ}+4\lambda^{0a}_{[I} x_{J]}\approx 0.
\ee
It can be split into two parts: the first gives a condition on the Lagrange multipliers
\be
\lambda^{0a}_I=-\hf\,\eps^{abc} F_{bc}^{IJ} x_J,
\label{cond-lambd0}
\ee
and the second is a primary constraint
\be
\CC^{a}_{I} = \eps_{IJKL}x^J\eps^{abc} F_{bc}^{KL}\approx 0.
\label{constrCC}
\ee
Taking into account the condition \eqref{B0-x},
the Hamiltonian is given by a linear combination of the primary constraints introduced above
\be
-H =\hf\,{\eps^{IJ}}_{KL} x_J B_{0a}^{KL} \CC^a_{I}+\omega_0^{IJ} \CG_{IJ}+\eps_{abc}\lambda^{ab}_I\Phi^{c}_I.
\label{Hamilgauge}
\ee

Now one should find the conditions under which the primary constraints are preserved in time.
Since their evolution is generated by the Hamiltonian \eqref{Hamilgauge}, this boils down
to the study of the constraint algebra. Introducing the smeared constraints
\be
\CG(n)=\int \de^3 \xc\, n^{IJ}\CG_{IJ},
\qquad
\CC^a(v)=\int \de^3\xc\, v^{I} \CC^{a}_{I},
\ee
the non-trivial commutators are given by
\be
\begin{split}
\{\CG(n),\CG(m)\}&\,=\CG([n,m]),
\\
\{\CG(n),\CC^a(v)\}&\,= \CC^a(n\cdot v)-4\int \de^3 \xc\,{\eps^{IJ}}_{KL}\lambda^{0a}_I x_J v^K n^{LM} x_M,
\\
\{\CG(n),\Phi^a_I\}&\,= -n^{IJ}\Phi^a_J+\tP^a_{IJ} n^{JK}x_K ,
\\
\{ \CC^a(v),\Phi^b_I \}&\,=-2\eps_{IJKL}v^J x^K\eps^{abc}D_c x^L,
\end{split}
\label{commprimary}
\ee
where we used \eqref{cond-lambd0}.
To apply these results, we start from the primary constraint $\Phi^a_I$. Its conservation in time leads to the condition
\be
\dot \Phi^a_I =
2\eps^{abc} B_{0b}^{IJ}D_c x_J+\tP^a_{IJ}\(\omega_0^{JK}x_K+\dot x^J\)\approx 0.
\ee
Combined with the three equations following from the Gauss constraint
\be
x^J\CG_{IJ}=D_a\Phi^a_I-\tP^a_{IJ} D_a x^J \approx 0,
\label{constr-on-Gauss}
\ee
it gives rise to the condition on the Lagrange multipliers $\omega_0^{IJ}$
\be
D_0 x^I=0
\label{condom0}
\ee
and to the secondary constraints
\be
\Psi_a^I=D_a x^I\approx 0.
\label{secondarygauge}
\ee
Then the conservation of the Gauss constraint $\CG_{IJ}$ gives a relation between the Lagrange multipliers $\lambda^{\mu\nu}_I$
\be
\dot\CG_{IJ}\approx -4x^{[I}B_{0a}^{J]K}\lambda^{0a}_K-\eps_{abc}x_{[I}\tP^a_{J]K}\lambda^{bc}_K=0
\quad \Longrightarrow \quad
B_{\mu\nu}^{IJ}\lambda^{\mu\nu}_J=0,
\label{partcondlambda}
\ee
whereas the conservation of $\CC^a_I$ does not generate new conditions.

The next step is to study the secondary constraints $\Psi^a_I$ \eqref{secondarygauge}.
In fact, due to the identity $x_I \Psi_a^I=0$, they give only 9 independent equations.
These constraints satisfy the following commutation relations
\be
\{ \CC^a(v),\Psi_b^I\}=0,
\qquad
\{ \CG(n),\Psi_b^I\}=x_J D_b n^{IJ},
\qquad
\{\Phi^a_I,\Psi_b^J\}=-\hf\,\delta^a_b(\delta_I^J-x_I x^J)\delta(\xc,\yc).
\label{commPsi}
\ee
Taking into account the restriction \eqref{condom0}, the conservation of $\Psi_a^I$ therefore
amounts to vanishing of all Lagrange multipliers $\lambda^{ab}_I=0$.
(Recall that these multipliers can be chosen to satisfy $x^I\lambda^{ab}_I=0$
from the very beginning so that the number of their
independent components equals the number of independent secondary constraints.)
Moreover, it is easy to check that
\be
\eps^{abc} F_{bc}^{IJ}x_J =2\eps^{abc}D_b\Psi_c^I
\qquad \Longrightarrow \qquad
\lambda^{0a}_I=0,
\ee
which fixes the last undetermined part of the Lagrange multipliers $\lambda^{\mu\nu}_I$.
Thus, the stabilization procedure stops at this point.

Once we found all the constraints, we can study whether they are of first or second class.
Due to \eqref{constr-on-Gauss}, we remain only with four types of constraints:
\be
\CC^a_I,
\qquad
\hat \CG_I\equiv{\eps_{IJ}}^{KL}x^J \CG_{KL},
\qquad
\Phi^a_I,
\qquad
\Psi_a^I.
\label{rem-constr}
\ee
Furthermore, due to the Bianchi identity, the first ones additionally satisfy
\be
D_a\CC^a_I=2\Psi_a^J \CC^a_{[I} x_{J]}+2{\eps_{IJKL}}x^J\eps^{abc}\Psi_a^K D_b\Psi_c^{L}.
\ee
Using the commutation relations \eqref{commprimary} and \eqref{commPsi}, it is trivial to check
that the constraints $\CC^a_I$ and $\hat \CG_I$ are first class, whereas
$\Phi^a_I$ and $\Psi_a^I$ are second class.
As a result, the $18+18=36$ configuration variables are restricted by $(9-3)+3=9$ first class and $9+9=18$
second class constraints, which leaves us with a zero dimensional physical phase space.
This confirms that this theory is topological, i.e. it does not contain propagating degrees of freedom.

\subsection{Reduction to SU(2) BF theory}
\label{subsec-reduct}

The meaning of the constraints \eqref{rem-constr} is quite transparent: the second class constraints,
$\Phi^a_I$ and $\Psi_a^I$, fix the off-diagonal (boost) degrees of freedom in the chiral decomposition of the $\so(4)$ algebra
or, more precisely, the configuration variables from the orthogonal completion to the $\su(2)_x$ subalgebra.
The remaining constraints, which describe the dynamics of the variables
from this subalgebra, are nothing else but the usual constraints of the $\SU(2)$ BF theory \cite{Mondragon:2004nw}.
Namely, $\CC^a_I$ gives the flatness condition for an $\SU(2)$-connection and $\hat \CG_I$ is the corresponding
Gauss constraint generating $\SU(2)$ gauge transformations.
This becomes especially clear in the time gauge $x^I=\delta^I_0$ where the second class constraints simply mean
that $\tPp{a}{}=\tPm{a}{}$ and $\omega_a^{(+)}=\omega_a^{(-)}$.
For a constant $x^I$ these relations get rotated by a constant $\Sp(4)$ transformation mapping $\delta^I_0$ into $x^I$.
It is however amusing to see how they generalize to the case of arbitrary gauge where $x^I$ can vary in spacetime.

Let us introduce the projections of our fields on the $\su(2)_x$ subalgebra
\be
\tp^a_{IJ}=I_{IJ}^{KL}(x)\tP^a_{KL},
\qquad
b_{0a}^{IJ}=I^{IJ}_{KL}(x)B_{0a}^{KL},
\ee
where $I_{IJ}^{KL}(x)$ is the projector given in \eqref{proj},
so that the new fields solve \eqref{Phi-x} and \eqref{B0-x}, respectively. Besides, we define
the following connection
\be
A_\mu^{IJ}=I^{IJ}_{KL}(x)\omega_\mu^{IJ}+2x^{[I}\p_\mu x^{J]},
\label{conA}
\ee
where the last term takes care about variations of the normal field.
This connection coincides with the original spin connection $\omega_\mu^{IJ}$
on the surface of \eqref{condom0} and \eqref{secondarygauge}
and satisfies the constraint
\be
J^{IJ}_{KL}(x)A_\mu^{KL}=2x^{[I}\p_\mu x^{J]},
\label{conSAA}
\ee
where $J(x)$ is the projector orthogonal to $I(x)$ (see appendix \ref{ap-algernra}).
This constraint is identical to the one appearing in the Lorentz covariant formulation of loop quantum gravity
\cite{Alexandrov:2001wt,Alexandrov:2002br}.
The characteristic feature of the connection satisfying \eqref{conSAA}
is that its holonomies map a vector from $\su(2)_{x_{1}}$ to
a vector in $\su(2)_{x_{2}}$ subalgebra \cite{Alexandrov:2010pg}, i.e. for constant $x^I$ they belong
to the $\SU(2)_x$ subgroup.

In terms of the new variables and taking into account all conditions on the Lagrange multipliers,
the 3+1 decomposed action \eqref{actiongauge} can be written as
\be
\begin{split}
S_\text{deg-gr}=&\,\int_{\Rmat}\de t\int_\Sigma \de^3\xc\Bigl[
\tp^a_{IJ}\p_0 A_a^{IJ}+2\Phi^a_I\p_0 \Psi_a^I +\eps^{abc} b_{0a}^{IJ} \(F_{bc}^{IJ}(A)-2\Psi_b^I\Psi_c^J\)
\Bigr.\\
&\, \Bigl.
+A_0^{IJ}\(\CG^{(A)}_{IJ}+2\Phi^a_I\Psi_a^J\)
\Bigr],
\end{split}
\label{act-decomp}
\ee
where $F_{ab}^{IJ}(A)$ is the curvature of the connection $A_a^{IJ}$ and
\be
\CG^{(A)}_{IJ}=\p_a\tp^a_{IJ}+[A_a,\tp^a]^{IJ}.
\ee
Setting the second class constraints to zero, one obtains the usual BF action
where all variables are projected down to $\SU(2)_x$.
It provides a covariant embedding of the $\SU(2)$ BF theory
into the $\Sp(4)$ formalism.

\subsection{Summary}
\label{subsec-sum}

Let us summarize what we have found.
The theory \eqref{plebanskiaction2} describing the degenerate sector of Plebanski formulation of general relativity
has two sectors of solutions of the simplicity constraints \eqref{simplicity constraint}. In the sector which
we called ``degenerate gravitational'', it reduces to the four-dimensional $\SU(2)$ BF theory covariantly
embedded into $\Sp(4)$ gauge group. The embedding is characterized by the normal vector $x^I$.
In the partially fixed gauge where $x^I$ is a fixed function, the theory possesses two types of second class constraints
conjugate to each other:
\be
\text{primary}\quad \Phi^{a}_{I}=x^J \tP^a_{IJ},
\qquad
\text{secondary}\quad \Psi_a^I=D_a x^I.
\label{final-constr}
\ee
The remaining constraints are first class and generate the gauge symmetries of the $\SU(2)$ BF theory.

The presence of the second class constraints, as usual, leads to a modification of the symplectic structure:
the Poisson bracket \eqref{Poisbr} has to be replaced by the appropriate Dirac bracket.
The latter can be easily calculated and is given by
\be
\{\omega_a^{IJ}(\xc), \tP^b_{KL}(\yc)\}_D=\delta_a^b I^{IJ}_{KL}(x)\delta(\xc,\yc)
\label{Dirbr}
\ee
with all other commutators being vanishing.
This result again demonstrates that only the $\SU(2)_x$ part of the configuration variables
is dynamical and implies that $\tP^a_{IJ}$ and $\omega_a^{IJ}$ are not canonically conjugate anymore.

\subsection{Inclusion of the Immirzi parameter}
\label{subsec-impar}

It is easy to include the Immirzi parameter \cite{Immirzi:1996dr} into our model.
To this end, one makes the usual replacement of the $B$-field in the BF part of the action
by the combination $B+\frac{1}{\gamma}\star B$. It leads to a mixing of the two solution sectors \eqref{twosectors}.
Now both of them reduce to the $\SU(2)$ BF theory,
just in the deg-topological case the resulting action is multiplied by the factor $1/\gamma$.

In the Hamiltonian formulation of the deg-gravitational sector described by the action
\be
\label{actiongauge-im}
S_\text{deg-gr}^{(\gamma)}[\omega,B,\lambda;x]=\hf\int_\CM\de^4\xc
\left[\eps^{\mu\nu\rho\sigma}\(B_{\mu\nu}^{IJ}+\frac{1}{2\gamma}\,{\eps^{IJ}}_{KL}B_{\mu\nu}^{KL}\)F_{\rho\sigma}^{IJ}
+4\lambda^{\mu\nu}_I x_J B_{\mu\nu}^{IJ}\right],
\ee
the presence of the Immirzi parameter affects the Poisson symplectic structure
so that the new canonical variables are
\be
\omega_{a}^{IJ}
\qquad
{\rm and}
\qquad
\tPb{a}{IJ}=\(1+{\gamma}^{-1}\star\)\tP^a_{IJ}.
\label{dynvar-im}
\ee
Nevertheless, all the remaining structure does not change and there are just slight modifications
in the stabilization procedure of section \ref{subsec-canan}.
In particular, the final constraints acting on the phase space
\eqref{dynvar-im} comprise $\CC^a_I$ and $\hat\CG_I$ which are first class
and $\Phi^a_I$, $\Psi_a^I$ which are second class, and all these constraints are given by the same expressions as
without $\gamma$. Moreover, due to the second class constraints, the $\gamma$-dependent symplectic structure given by
Poisson brackets is replaced by the $\gamma$-independent one described by the same Dirac brackets \eqref{Dirbr} as before.

\section{Spin foam quantization}
\label{sec-quant}

In this section we consider the spin foam quantization of our model.
Since after implementing the second class constraints (in the deg-gravitational sector)
it coincides with the $\SU(2)$ BF theory, we know what the final result should be:
it is given by the Crane-Yetter model \cite{Ooguri:1992eb,Crane:1993if}
with the structure group $\SU(2)$ represented by the following spin foam state sum
\be
\mathcal{Z}_{\rm CY}^{\SU(2)}(\Delta^*)
=\sum_{j\rightarrow f}\sum_{\inter\rightarrow e}\prod_{f\in\Delta^*}(2j_f+1)\prod_{v\in\Delta^*}A_v^{\SU(2)},
\label{pfCY}
\ee
where $\Delta^*$ is a two-complex dual to a simplicial triangulation $\Delta$ of the spacetime manifold $\CM$,
$j_f$ labels irreducible representations of $\SU(2)$ attached to the faces of $\Delta^*$,
$\inter_e$ are $\SU(2)$ invariant intertwiners assigned to its edges,
and $A_v^{\SU(2)}$ is the vertex amplitude given by the $\SU(2)$ $\{15j\}$ symbol.
The latter is obtained by evaluation of the $\SU(2)$ spin network represented by the pentagon graph,
which is dual to the boundary of a 4-simplex $\sigma\in\Delta$ dual to the vertex $v\in \Delta^*$,
\ba
~\nonumber\\
~\nonumber\\
~\label{vampBF}
\ea
\begin{center}
\unitlength 0.3mm
\linethickness{0.2pt}
\ifx\plotpoint\undefined\newsavebox{\plotpoint}\fi 
\begin{picture}(120,100)(0,0)
\put(-40,111.25){\makebox(0,0)[cc]{$A_v^{SU(2)}(\vec \jmath,\vec \imath)\ =\{15j\}\equiv$}}
\put(90,35){\circle*{6}}
\put(190,35){\circle*{6}}
\put(60,110){\circle*{6}}
\put(220,110){\circle*{6}}
\put(140,165){\circle*{6}}
\put(60,110){\line(1,0){160}}
\multiput(60,110)(.04904966278,.03372164316){1631}{\line(1,0){.04904966278}}
\multiput(140,165)(.04904966278,-.03372164316){1631}{\line(1,0){.04904966278}}
\put(220,110){\line(-2,-5){30}}
\put(190,35){\line(-1,0){100}}
\put(90,35){\line(-2,5){30}}
\multiput(60,110)(.058479532164,-.033738191633){2223}{\line(1,0){.058479532164}}
\multiput(190,35)(-.03373819163,.08771929825){1482}{\line(0,1){.08771929825}}
\multiput(140,165)(-.03373819163,-.08771929825){1482}{\line(0,-1){.08771929825}}
\multiput(90,35)(.058479532164,.033738191633){2223}{\line(1,0){.058479532164}}
\put(78.5,29){\makebox(0,0)[cc]{$\inter_{1}$}}
\put(47,111.25){\makebox(0,0)[cc]{$\inter_{2}$}}
\put(145.25,172.5){\makebox(0,0)[cc]{$\inter_{3}$}}
\put(235,111.25){\makebox(0,0)[cc]{$\inter_{4}$}}
\put(204.75,29){\makebox(0,0)[cc]{$\inter_{5}$}}
\put(64.25,69){\makebox(0,0)[cc]{$j_{12}$}}
\put(88.75,143.5){\makebox(0,0)[cc]{$j_{23}$}}
\put(191.5,143.5){\makebox(0,0)[cc]{$j_{34}$}}
\put(216.25,69){\makebox(0,0)[cc]{$j_{45}$}}
\put(140,26.5){\makebox(0,0)[cc]{$j_{51}$}}
\put(140,117.75){\makebox(0,0)[cc]{$j_{24}$}}
\put(127.25,82.5){\makebox(0,0)[cc]{$j_{25}$}}
\put(153.25,82.5){\makebox(0,0)[cc]{$j_{14}$}}
\put(104,98){\makebox(0,0)[cc]{$j_{13}$}}
\put(176.75,98){\makebox(0,0)[cc]{$j_{53}$}}
\put(-30,105){\makebox(0,0)[cc]{}}
\end{picture}
\end{center}
\vspace{-0.6cm}
The normalization of the intertwiners used in this evaluation is defined in appendix \ref{ap-ClG}.

However, we would like to proceed in a different way which would avoid quantizing the degrees
of freedom on the constraint surface only. Our aim is to find a quantization of the original
theory with constraints, \eqref{plebanskiaction2} or \eqref{actiongauge} (or even \eqref{actiongauge-im}), such that
it reproduces the $\SU(2)$ Crane-Yetter model \eqref{pfCY}. In particular, we would like to check whether
the quantization strategies used to get the EPRL or the FK model are able to do this.
Since our model is only slightly different from the gravitational sector of Plebanski theory,
our study represents a very serious test on the validity of these quantization approaches.

\subsection{Discretization}
\label{subsec-discr}

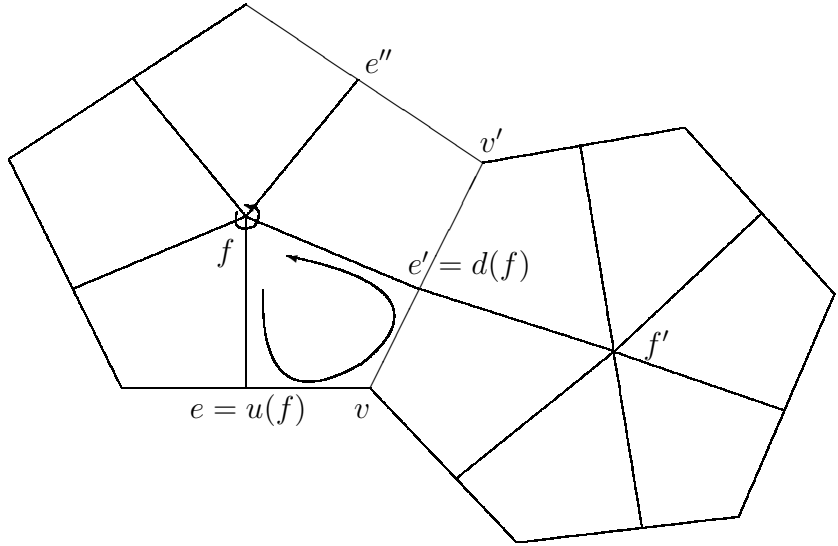
\begin{figure}
\unitlength .6mm 
\linethickness{0.4pt}
\ifx\plotpoint\undefined\newsavebox{\plotpoint}\fi 
\begin{picture}(258,125)(-30,10)
\put(70,50){\line(1,0){55}}
\put(125,50){\line(1,2){25}}
\put(150,100){\line(-3,2){52.5}}
\multiput(97.5,135)(-.1033464567,-.0674212598){508}{\line(-1,0){.1033464567}}
\multiput(45,100.75)(.0673854447,-.1367924528){371}{\line(0,-1){.1367924528}}
\put(97.5,50){\line(0,1){38}}
\multiput(97.5,88)(-.160714286,-.067226891){238}{\line(-1,0){.160714286}}
\multiput(135.75,72)(-.160714286,.067226891){238}{\line(-1,0){.160714286}}
\multiput(97.5,88)(.0673854447,.0822102426){371}{\line(0,1){.0822102426}}
\multiput(97.5,88)(-.0673854447,.0822102426){371}{\line(0,1){.0822102426}}
\multiput(125,50)(.0674273859,-.0710580913){482}{\line(0,-1){.0710580913}}
\multiput(157.5,15.75)(.57267442,.06686047){86}{\line(1,0){.57267442}}
\multiput(206.75,21.5)(.0674603175,.1571428571){315}{\line(0,1){.1571428571}}
\multiput(150,100)(.389130435,.067391304){115}{\line(1,0){.389130435}}
\multiput(194.75,107.75)(.0674442191,-.0750507099){493}{\line(0,-1){.0750507099}}
\multiput(179,58.25)(-.210784314,.067401961){204}{\line(-1,0){.210784314}}
\multiput(179,58)(-.0837349398,-.0674698795){415}{\line(-1,0){.0837349398}}
\multiput(179,58)(.067204301,-.416666667){93}{\line(0,-1){.416666667}}
\multiput(179,58)(.195595855,-.067357513){193}{\line(1,0){.195595855}}
\multiput(179,58)(.0718201754,.0674342105){456}{\line(1,0){.0718201754}}
\multiput(179,58)(-.066964286,.410714286){112}{\line(0,1){.410714286}}
\put(98,45){\makebox(0,0)[cc]{$e=u(f)$}}
\put(123.25,45){\makebox(0,0)[cc]{$v$}}
\put(93,80){\makebox(0,0)[cc]{$f$}}
\put(188.5,59.5){\makebox(0,0)[cc]{$f'$}}
\put(147.,76.95){\makebox(0,0)[cc]{$e'=d(f)$}}
\put(152,105){\makebox(0,0)[cc]{$v'$}}
\put(127,122.5){\makebox(0,0)[cc]{$e''$}}
\qbezier(101.25,71.75)(100.5,42.5)(122.75,55.25)
\put(106.75,79.25){\vector(-4,1){.141}}\qbezier(123,55.25)(143.625,71.75)(106.75,79.25)
\put(97.25,90.5){\vector(-4,1){.141}}\qbezier(99.25,85.75)(102.25,90.375)(97.25,90.5)
\qbezier(99.25,85.75)(94.125,84.125)(95.5,89)
\end{picture}
\caption{Two faces sharing edge $e'$ and divided into wedges. The small arrow indicates the orientation of the face and
the big arrow shows how one evaluates the curvature associated to the wedge. For a given vertex $v$, the two adjacent
edges $e$ and $e'$ are denoted as $u(f)$ and $d(f)$, according to the orientation of $f$ to which they both belong.}
\label{fig-wedge}
\end{figure}

Before we start discussing the quantization, let us discretize the variables of our model given
a simplicial decomposition $\Delta$ of the spacetime manifold.
As usual, the $B$-field is discretized by associating Lie algebra valued elements $B_f\in\so(4)$
to the faces of the dual two-complex $\Delta^*$,
which can be obtained as integrals of the $B$-field over the dual triangles\footnote{More precisely,
to make sense of the integral, the $B$-field at different points in \eqref{defbiv} should be parallel
transported to a reference point. This introduces a dependence of the bivectors $B_f$ on the connection, which explains
their mutual non-commutativity with respect to the Poisson symplectic structure \cite{Ashtekar:1998ak,Baratin:2010nn}.}
\be
B_f^{IJ}=\int_{t_f} B^{IJ}.
\label{defbiv}
\ee
The spin connection $\omega^{IJ}$ gives rise to group elements $g_e$
which coincide with its holonomies along the edges $e\in\Delta^*$.
However, it is also convenient to introduce the holonomies $g_{ve}$, going from
vertex $v$ to the center of edge $e$,
and $g_{fe}$, going from the center of face $f$ also to the center of the edge.
$g_{ev}$ and $g_{ef}$ will denote their inverse.
The former provide a refined version of our basic dynamical variables
which are obtained as $g_e=g_{ve}g_{ev'}$,
where $v$ and $v'$ are the two vertices joined by $e$.
On the other hand, $g_{fe}$ are needed to bring the bivectors \eqref{defbiv}
to the reference frame where $g_{ve}$ are acting. In particular, we define
\be
B_{ef}=g_{ef}B_f g_{fe}.
\label{defBfe}
\ee
In this sense, these group elements may be considered as non-dynamical auxiliary variables
completing the definition of the discrete $B$-field.
Altogether, $g_{ve}$ and $g_{fe}$ provide the discretization of the spin connection
on the two-complex obtained by subdivision of $\Delta^*$ into wedges, which are in one-to-one
correspondence with pairs $(vf)$, as shown on Fig. \ref{fig-wedge} \cite{Reisenberger:1997sk}.

Finally, we should discretize the normal field $x^I$ appearing explicitly in the gauge fixed action \eqref{actiongauge}
and in the linearized simplicity constraints \eqref{twosectorslin}. This field is analogous to a similar field
appearing in the canonical formulation of the usual Holst and Plebanski actions \cite{Alexandrov:2000jw} where it describes
the normal to the three-dimensional spacelike slices. In spin foam models it usually gives rise to the normal vectors $x_e$,
which can be viewed as elements of the factor space $X=\Sp(4)/\SU(2)$
\cite{Alexandrov:2007pq,Alexandrov:2008da,Gielen:2010cu,Baratin:2011hp}.
Such a vector is interpreted geometrically as the normal to the tetrahedron dual to edge $e$.
However, in the degenerate case it is more natural to associate such normal vectors to 4-simplices.
Thus, at the discrete level the field $x^I$ will be represented by a set of unit vectors $x_v$.

With these definitions we can now discretize the simplicity constraints.
Their quadratic form \eqref{simplicity constraint}
is discretized as usual giving rise to diagonal, cross and volume simplicity
obtained by averaging the two bivectors over the same triangle, different triangles belonging
to the same tetrahedron, or non-intersecting triangles of the same 4-simplex, respectively
\cite{DePietri:1998mb,Freidel:1999rr}.
The only difference with the usual case is the form of the volume constraint which requires that the geometric volume
of the 4-simplex vanishes.

On the other hand, the discretization of the linear simplicity constraints \eqref{gravsectorlin}
involves an extra ingredient. Indeed, they relate the fields, $x^I$ and $B^{IJ}$, which after discretization
live at different elements of the cellular decomposition: at vertices and faces, respectively.
Due to this, the bivectors should be transported to the reference frame of a vertex using the holonomies
introduced above. As a result, the discrete simplicity constraints read as follows
\be
g_{ve} B_{ef} g_{ev} \cdot x_v=0,
 \quad \text{for }\forall f \supset e\supset v.
\label{lin-simple-discr}
\ee
It is important to notice that since there are two ways to connect a face to a vertex (by going either
through $e=u(f)$ or $e=d(f)$), the conditions \eqref{lin-simple-discr} constrain not only the bivectors, but
also the holonomies. This shows that at the discrete level
the primary and secondary constraints are not well distinguished from each other.

To make contact with the new spin foam models, it is convenient to change a bit the point of view
and to write the simplicity constraints in the reference frame of a tetrahedron or its dual edge.
To this end, we define
\be
x_e(v)=g_{ev}x_v
\label{defxev}
\ee
so that the condition \eqref{lin-simple-discr} becomes
\be
B_{ef}^{IJ}(x_e(v))_J=0,\quad \text{for }\forall f\supset e\supset v.
\label{discr-simple-edge}
\ee
Up to insertion of the Hodge operator,
this is the usual form of the linear simplicity constraints used in the new spin foam models.
In our case it should be supplemented by the additional requirement
that the normals $x_e(v)$ originate from the same vector $x_v$
and therefore must satisfy
\be
g_{ve}x_e(v)=g_{ve'}x_{e'}(v), \quad \text{for }\forall e,e'\supset v.
\label{cond-vef}
\ee

Then we turn to the secondary second class constraints \eqref{secondarygauge}.
They restrict the holonomies of the spin connection
and at the discrete level read as
\be
x_v=g_{e}x_v', \quad \text{for } e\supset v,v'
\label{xvxe}
\ee
Being combined with \eqref{defxev}, they can be equivalently rewritten as
\be
x_e(v)=x_e(v').
\label{xevxev}
\ee
This shows that, provided the secondary constraints are imposed, one can drop the $v$-dependence
of the normals $x_e$. Moreover, this suggests to introduce the normal vectors associated
to all elements of the spin foam cellular complex: $x_v$, $x_e$ and $x_f$.
This pluralistic point of view allows to formulate all constraints in a simple and uniform way.
Indeed, they become equivalent to the following relations
\begin{subequations}
\beq
& B_{f}^{IJ}(x_f)_J=0, &
\label{discr-simple}
\\
& x_e=g_{ev}x_v,
\qquad
x_f=g_{fe}x_e. &
\label{xfxe}
\eeq
\label{alldiscrconstr}
\end{subequations}
These relations represent the straightforward discretization of the primary and secondary second class constraints
\eqref{final-constr}, respectively. We emphasize that it is crucial to consider
the primary and secondary constraints on equal footing.
For example, taken alone, \eqref{discr-simple} is not sufficient to generate
the degeneracy condition for a 4-simplex. On the other hand, altogether the conditions \eqref{alldiscrconstr}
provide an elegant and amazingly simple discrete formulation of all second class constraints
of the continuous theory.

To facilitate the use of the constraints in the discretized path integral, let us rewrite them using
the chiral decomposition of $\Sp(4)$ (see appendix \ref{ap-algernra}).
To this end, we note that each normal vector $x\in X=\Sp(4)/\SU(2)$ gives rise to an element $\xb\in \SU(2)$ defined by
\be
\xb=g_x^-(g_x^+)^{-1},
\label{def-uu}
\ee
where $g_x\in\Sp(4)$ is a representative of $x$ which choice does not affect the definition of $\xb$.
This definition implies that $g\cdot x$ is mapped to $g^-\xb (g^+)^{-1} $.
Then it is easy to see that \eqref{alldiscrconstr} is equivalent to\footnote{To arrive at \eqref{discr-simple-group},
it is useful to note that \eqref{discr-simple} is equivalent to the statement that
the left and right components of $g_{x_f}^{-1}B_f g_{x_f}$ are equal. Then \eqref{discr-simple-group}
follows by applying the definition of $\xb_f$ \eqref{def-uu}.}
\begin{subequations}
\beq
& B_f^-=\xb_f B_f^+ \xb_f^{-1}, &
\label{discr-simple-group}
\\
& g_{ve}^-=\xb_v g_{ve}^+ \xb_e^{-1},
\qquad
g_{fe}^-=\xb_f g_{fe}^+ \xb_e^{-1}. &
\label{xfxe-group}
\eeq
\label{chiral-constr}
\end{subequations}
The other constraints appearing above like \eqref{discr-simple-edge} and \eqref{cond-vef} have a similar representation.

\subsection{The usual strategy}
\label{subsec-usual}

Our first aim is to apply to our model the quantization strategy employed in the EPRL and FK spin foam models
and based on the idea ``first quantize, then constrain".
This implies that one should start from the unconstrained $\Sp(4)$ BF theory and
incorporate the simplicity constraints at quantum level.
The spin foam quantization of the unconstrained theory is provided by the Crane-Yetter model \cite{Crane:1993if}
with the structure group $\Sp(4)$ represented by the following state sum
\be
\mathcal{Z}_{\rm CY}^{\Sp(4)}(\Delta^*)
=\sum_{\lambda=(j^+,j^-)\rightarrow f}\sum_{\Int=(\inter^+,\inter^-)\rightarrow e}
\prod_{f\in\Delta^*}(2j^+_f+1)(2j^-_f+1)\prod_{v\in\Delta^*}\{15j^+\}\{15j^-\},
\label{pfSpin4}
\ee
where the sum goes over all $\Sp(4)$ irreducible representations $\lambda_f$ and all $\Sp(4)$ invariant intertwiners $\Int_e$.
The simplicity constraints in this approach are supposed to restrict the allowed set of representations and intertwiners
such that the resulting state sum provides the discretized path integral for the constrained theory.
In our case, the result must reproduce the partition function \eqref{pfCY}.

However, it is clear that if the effect of the constraint imposition is only the reduction of
the admissible group theoretic data, the state sum \eqref{pfSpin4} will never reduce to \eqref{pfCY}.
Indeed, the two partition functions have different vertex amplitudes and it is impossible to reduce one to another
by restricting the kinematical data.

This approach can be realized explicitly by proceeding as follows.
As is well known, the asymptotic analysis of both EPRL and FK models reveals
that they contain a degenerate sector \cite{Conrady:2008mk,Barrett:2009gg}.
Its geometric interpretation precisely corresponds to the classical geometries described by our model.
Thus, a simple way to get a spin foam quantization of \eqref{actiongauge-im}
is to extract the degenerate sector from the EPRL or FK state sum.
This can be achieved by expressing this state sum as an integral over coherent states
\cite{Freidel:2007py,Livine:2007ya}. It has been shown \cite{Barrett:2009gg} that each quasiclassical Regge geometry
contributing to the asymptotics of this integral can be uniquely reconstructed from the boundary data
consisting of the $SU(2)$ representations associated to triangles of $\Delta$ (or faces of $\Delta^*)$
and coherent states assigned to each pair $(ef)$.
The degenerate sector is then extracted by restricting only to those boundary data which lead to degenerate
Regge geometries.
Since this restriction is imposed only on the boundary data, it affects only the set of
representations and intertwiners which survive in the partition function \eqref{pfSpin4}
after imposing the simplicity constraints. It does not affect the general form of the vertex
amplitude and therefore cannot reproduce the desired result \eqref{vampBF}.

This method of imposing the degeneracy condition has a clear drawback:
it ensures the vanishing of the geometric volume of 4-simplices only in the quasiclassical limit.
On the other hand. this condition appears as a part of our simplicity constraints
which are expected to hold at the full quantum level.
This suggests that one should look for an alternative approach. Such an approach does exist
and moreover it leads to the correct vertex amplitude.
But to realize it, we should be ready to go beyond the usual strategy and to constrain
not only the kinematical data in \eqref{pfSpin4}, but also the group elements entering the definition of the vertex amplitude.

In the previous section we showed that the simplicity constraints can be represented as a combination of
two conditions, \eqref{discr-simple-edge} and \eqref{cond-vef}. The former are the usual linear simplicity constraints
in the non-degenerate sector of Plebanski formulation. The new spin foam models provide a way to implement them
at quantum level so that this step can be considered as being already accomplished.
Thus, it remains only to incorporate the second condition, which restricts us precisely
to the degenerate sector by requiring that the normals to all tetrahedra of a 4-simplex,
transferred to the same frame, coincide.

Rather than a constraint on the normals $x_e(v)$, the condition \eqref{cond-vef} can be viewed as
a constraint on the holonomies $g_{ve}$. These holonomies coincide with the group elements appearing in
the integral formula for the vertex amplitude of the new models
\be
A_v^{(\gamma)}(j_f,k_{e f},\inter_e)=\int \prod_{e\supset v}\de \gl_{ve}\,
\CS_{(\Gamma_v,\lambda^{(\gamma)}(j_f),k_{e f},\inter_e)}\[\gl_{vu(f)}^{-1}\gl_{vd(f)}^{\mathstrut},x_e(v)\],
\label{gam-amplit_simplex}
\ee
where $u(f)$ and $d(f)$ denote the two edges belonging to the face $f$ and sharing the vertex $v$
(one of them is considered as ``up" and the other as ``down", see Fig. \ref{fig-wedge}),
and $\CS_{(\Gamma_v,\lambda_f,k_{e f},\inter_e)}\[\gf_f,x_e\]$ is the so-called projected spin network
\cite{Livine:2002ak} defined on the graph $\Gamma_v$ dual to the boundary of a 4-simplex
(this is the same graph as the one appearing in \eqref{vampBF}). The projected spin network
is labeled by $\Sp(4)$ representations $\lambda_f$
attached to the links of the graph, $\SU(2)$ representations $k_{ef}$ assigned to the ends of the links, and
$\SU(2)$ invariant intertwiners $\inter_e$ associated to the nodes.
In both EPRL and FK models, the imposition of the simplicity constraints \eqref{discr-simple-edge}
leads to that the $\Sp(4)$ representations $\lambda_f$ are defined in terms of
the SU(2) representations $j_f$ as\footnote{One should take into account that the constraints \eqref{discr-simple-edge}
differ from the linear simplicity constraints
describing the gravitational sector in the new spin foam models by the absence of the Hodge operator, i.e.
they are actually analogous to the constraints specifying the topological sector (see footnote \ref{foot-twoconstr}).
This difference can be accounted by replacing the Immirzi parameter by its inverse.}
\be
\lambda^{(\gamma)}(j)=\(\frac{1}{2\gamma}\,(1+\gamma)j,\frac{1}{2\gamma}\,|1-\gamma|j\).
\label{lambdaj}
\ee
On the other hand, the representations $k_{ef}$ are treated differently:
in the FK model they can be arbitrary, whereas in the EPRL model they are fixed to be $k_{ef}=j_f$.
(For $\gamma=0$, the EPRL prescription gives $\lambda^{(0)}(j)=(j,j)$, $k_{ef}=0$ and reproduces
the BC model.)

Given the integral representation \eqref{gam-amplit_simplex}, the natural idea to incorporate the condition \eqref{cond-vef}
is to insert it into the measure. This amounts to adding the factor $\delta\((g_{ve}^-)^{-1}\xb_v g_{ve}^+ \xb_e^{-1}(v)\)$
and it is straightforward to evaluate the resulting vertex amplitude.
To this end, let us recall the explicit expression for the projected spin network
\be
\CS_{(\Gamma_v,\lambda_f,k_{e f},\inter_e)}\[\gf_f,x_e\]=
\bigotimes_{e\supset v} \inter_e \,\cdot\,
\bigotimes_{f\supset v} \ClG{j^+_f}{j^-_f}{k_{u(f)f}}{}{}{}
\Db^{(\lambda_f)}\(g_{x_{u(f)}}^{-1}\gf_f g_{x_{d(f)}}\)
\overline{\ClG{j^+_f}{j^-_f}{k_{d(f)f}}{}{}{}\vphantom{C^{A^{A^A}}}},
\label{form-prsn}
\ee
where $\Db^{(\lambda_f)}(g)$ is the image of $g\in\Sp(4)$ in
representation $\lambda_f=(j_f^+,j_f^-)$ and $\ClG{j_1}{j_2}{j_3}{}{}{}$ is the invariant map whose matrix elements are
given by the Clebsch-Gordan coefficients.
Then using the following property of the $\SU(2)$ matrix elements
\be
\mathop{\sum}\limits_{m,m',n,n'}\ClG{j^+}{j^-}{j_1^{\mathstrut}}{m^{\mathstrut}}{m'}{\ell_1}\,
\Db^{(j^+)}_{mn^{\mathstrut}}(h)\Db^{(j^-)}_{m'n'}(h) \,
\overline{\ClG{j^+}{j^-}{j_2}{n^{\mathstrut}}{n'}{\ell_2}}
=\delta_{j_1j_2} d_{j_1}^{-1}\, \Db^{(j_1)}_{\ell_1 \ell_2}(h),
\label{property}
\ee
where $d_j=2j+1$,
one can easily show that
\beq
A_v^{\rm (deg)}(j_f,k_{e f},\inter_e)
& \equiv & \int \prod_{e\supset v}\Bigl[\de \gl_{ve}^+\de \gl_{ve}^- \,\delta\((g_{ve}^-)^{-1}\xb_v g_{ve}^+ \xb_e^{-1}(v)\)\Bigr]
\CS_{(\Gamma_v,\lambda^{(\gamma)}(j_f),k_{e f},\inter_e)}\[\gl_{vu(f)}^{-1}\gl_{vd(f)}^{\mathstrut},x_e(v)\]
\nonumber \\
& =  &\{15j\}\prod_{f\supset v} d_{k_f}^{-1} \delta_{k_{u(f)f}k_{d(f)f}},
\label{gamma-amplit_simplex-model}
\eeq
where the $\{15j\}$ symbol is constructed out of ten representations $k_f=k_{u(f)f}=k_{d(f)f}$ and five
representations characterizing the intertwiners $\inter_e$, exactly as in \eqref{vampBF}.
Thus, up to a normalization factor, we reproduced the correct vertex amplitude of the $\SU(2)$
Crane-Yetter model \eqref{pfCY}!

This beautiful result seems to indicate that the quantization strategy realized by
the new spin foam models passes our consistency check. However, its close inspection raises several questions.
First of all, a striking feature of \eqref{gamma-amplit_simplex-model} is that
it does not depend at all on the restrictions on the representations obtained by imposing
the first part of the simplicity constraints \eqref{discr-simple-edge}: it holds for any set of $\lambda_f$ and $k_{ef}$.
On the other hand, these restrictions are at the core of the new spin foam models and
it is very puzzling that, after imposing the remaining part of the simplicity given by \eqref{cond-vef},
all information about them is completely erased.
The only case where some information remains corresponds to $\gamma=0$ in the EPRL model.
But it is even worse. In this case $k_f$ are fixed to be zero and we do not get the correct vertex at all.
Furthermore, although the final result \eqref{gamma-amplit_simplex-model}
is perfectly fine for all values of the Immirzi parameter,
the intermediate step \eqref{gam-amplit_simplex} is not defined for irrational $\gamma$.

All these issues suggest that the constraints put forward by the EPRL and FK models
are not really relevant for getting the correct spin foam dynamics, and even generate some strange effects
like the quantization of the Immirzi parameter. On the other hand, the correct dynamics is obtained
by imposing the constraints missing in the usual approach. In fact, there is a crucial difference between
\eqref{discr-simple-edge} and \eqref{cond-vef}: although they are both needed to discretize
the primary simplicity constraints, the latter are better seen as a discretization of the secondary constraints
\eqref{secondarygauge}. (Let us recall that at the discrete level there is no a clear distinction
between the two types of constraints such that exists in the continuum theory.)
Thus, the insertion of these constraints into the integration measure
is exactly what has been suggested in the introduction (see below \eqref{amplit_simplex-gen}),
in our previous works \cite{Alexandrov:2008da,Alexandrov:2010pg}, and in \cite{Geiller:2011kr}.
There it was claimed that the measure over holonomies should include
the delta function of the secondary second class constraints.
Here we see quite explicitly that this insertion is indeed necessary.

As a result, we arrive at the following situation. It is indeed possible to extract from the EPRL and FK models
the correct dynamics of the degenerate sector provided we incorporate the secondary second class constraints directly into
the definition of the vertex amplitude. However, this modification of the vertex
makes the imposition of the primary simplicity constraints completely irrelevant.
This questions the ability of these approaches to capture the right dynamics
in the gravitational sector where the secondary constraints have been ignored so far.

In fact, as we will show in the next subsections,
there is a consistent way of quantizing the theory \eqref{actiongauge-im}
which leads to the full correct result \eqref{pfCY},
including not only the vertex, but also the edge and face amplitudes.
It requires to take into account all constraints \eqref{alldiscrconstr}.
As a consequence, one can use any version of the primary simplicity constraints: either at vertices \eqref{lin-simple-discr},
or at edges \eqref{discr-simple-edge}, or even at faces \eqref{discr-simple} --- they all become equivalent.
However, these constraints become important only for gluing the contributions of different simplices, whereas
the vertex amplitude associated with a given simplex turns out to be completely determined by the secondary constraints
restricting holonomies.
These results strongly support the expectation that the usual strategy to the spin foam quantization, based
on the use of only the primary constraints, is not satisfactory.

\subsection{Canonically inspired quantization}
\label{subsec_lessons}

In this subsection we provide the rules to construct the partition function of
a constrained theory of Plebanski type. These rules summarize the results obtained for the vertex amplitude
in \cite{Alexandrov:2008da} and the quantization procedure for the three-dimensional model of \cite{Geiller:2011kr}
developed in \cite{Alexandrov:2011km}. Here we formulate them in a coherent way, which can be applied
in quite generic situations. In particular, in the next subsection these quantization rules
are applied to our model describing the degenerate sector of Plebanski theory, and shown
to reproduce the correct quantization given by the partition function \eqref{pfCY}.

We assume that the theory to be quantized has the structure of Plebanski formulation of general relativity,
i.e. it is represented as topological BF theory supplemented by primary constraints $\phi$, whose
time evolution generates secondary constraints $\psi$.
Furthermore, we impose the partial gauge fixing of the boost gauge freedom, as we did in section \ref{subsec-canan}.
Then the quantization procedure we propose involves the following steps:
\begin{enumerate}
\item
First, we need to discretize the primary and secondary second class constraints.
We assume that their discrete versions give certain restrictions
\be
\phi_{\text{discr}}(B,x)=0,
\qquad
\psi_{\text{discr}}(g,x;B)=\unit
\label{cosntr-discrete}
\ee
on the bivectors and the holonomies, respectively. Both constraints are expected to depend on the normals $x\in X$
assigned to the elements of the two-complex $\Delta^*$ and we allowed the secondary constraints to depend on the bivectors.
In the model we consider here this dependence will be absent, which significantly simplifies its spin foam representation.
However, it is expected to arise in the constraints describing the gravitational sector of Plebanski theory.

\item
Using the discrete constraints \eqref{cosntr-discrete}, we construct the measures
\be
\begin{split}
\CD^{(x)} [B]&\,=\Delta(B,x)\delta\(\phi_{\rm discr}(B,x)\)\de B,
\\
\CD^{(x;B)} [g]&\,=\delta\(\psi_{\rm discr}(\gl,x;B)\)\de g,
\end{split}
\label{measur}
\ee
where the first one includes the factor which represents a discretization of the determinant of the Dirac matrix,
$\left|\det\{\phi,\psi\}\right|$. Since typically it does not depend on the spin connection, it can be expressed
through the bivectors and the normals and therefore attributed to the first measure only.

\item
Given the measure for holonomies, we evaluate the following quantity, which we interpret as vertex amplitude,
\be
A_v(\lambda_f,k_{e f},\inter_e)=\int \prod_{e\supset v}\CD^{(x;B)} [\gl_{ve}]\,
\CS_{(\Gamma_v,\lambda_f,k_{e f},\inter_e)}\[\gl_{vu(f)}^{-1}\gl_{vd(f)}^{\mathstrut},x_e\].
\label{amplit_simplex}
\ee
This is the same formula as \eqref{amplit_simplex-gen} given in the introduction, where the boundary state
is taken to be the projected spin network $\CS_{(\Gamma_v,\lambda_f,k_{e f},\inter_e)}\[\gf_f,x_e\]$,
and generalizes \eqref{gam-amplit_simplex} and \eqref{gamma-amplit_simplex-model}.
In particular, it is important to emphasize that no restriction on $\lambda_f$ and $k_{ef}$ is assumed.
Due to the gauge invariance of projected spin networks
and to the following covariance property of the measure
\be
\CD^{(x;B)} \[\gl_{ve} \gb\]=\CD^{(\gb\cdot\, x;\,\gb B \gb^{-1})} [\gl_{ve}], \qquad \gb\in \Sp(4),
\label{trmeas}
\ee
the vertex amplitude \eqref{amplit_simplex} is independent of the normals. In contrast, if the measure depends on bivectors,
this dependence propagates to $A_v$ and, as a result, it cannot be viewed as a true vertex amplitude in the spin foam representation.
If however the measure is $B$-independent, as it happens in our model, the formula \eqref{amplit_simplex} does provide
the spin foam vertex.

\item
To achieve the correct gluing of the vertex contributions, one has to perform
several additional steps. The first of them is to evaluate what can be called
the vertex amplitude in the ``connection" representation
\be
A_v[\gf_f,x_e]=\sum_{\lambda_f,k_{e f},\inter_e}\(\prod_{f\supset v} d_{\lambda_f}\)\(\prod_{(e,f)\supset v} d_{k_{e f}}\)
A_v(\lambda_f,k_{e f},\inter_e)
\bar\CS_{(\Gamma_v,\lambda_f,k_{e f},\inter_e)}\[\gf_f,x_e\].
\label{reprampl}
\ee
In generic case this name is not quite precise because this quantity carries a dependence on the bivectors,
which we did not indicate explicitly, originating from the measure in the definition of $A_v(\lambda_f,k_{e f},\inter_e)$.

\item
The quantity \eqref{reprampl} can be already used to glue several vertex contributions.
However, such gluing can be written more elegantly if one first passes to the $B_f$-representation.
To this end, one defines
\be
A_v[\hat B_f,x_f]=\int \prod_{f\supset v}
\[ \de \gf_f\, \exp\left\{\I\Tr\((1+\gamma^{-1} \star)B_f\cdot g_{fu(f)}\gf_f g_{fd(f)}^{-1}\)\right\}\]A_v[\gf_f,x_e],
\label{amplitgen}
\ee
where $\gamma$ is the Immirzi parameter and we used the fact that the amplitude \eqref{amplitgen} depends
on bivectors only in the combination $\hat B_f\equiv g_{fd(f)}^{-1} B_f g_{fu(f)}$. The group elements $g_{fe}$
are restricted to satisfy the same type of the second class constraints $\psi_{\rm disc}$ as $g_{ve}$.
We do not integrate over them since they drop out from the final partition function. Alternatively, one could
insert integrals $\int \CD^{(x;B)}[g_{fe}]$ which explicitly put the constraints on these holonomies.
Note in contrast that the group elements $\gf_f$ are integrated with the standard Haar measure and not the one involving
the second class constraints.

\item
Finally, the total partition function is formed by multiplying the vertex amplitudes \eqref{amplitgen}
using the non-commutative star-product\footnote{This product is defined on the plane waves
$\eb_g(a)=e^{\I\Tr(g\cdot a)}$, with $a\in\gmat$
and momentum given by group element $g$, as  $\eb_{g_1}\star \eb_{g_2}=\eb_{g_1 g_2}$.}
\cite{Baratin:2010wi,Baratin:2010nn} and integrating the result over the bivectors with the measure \eqref{measur}
\be
\CZ=\int\prod_f  \CD^{(x)}[B_f] \(\,\mathop{\bigstar}\limits_v \,A_v[\hat B_f,x_f]\).
\label{fullpf-noncom}
\ee
The order in the non-commutative product is dictated by the orientation of faces and by the choice in each face
of a ``reference tetrahedron" \cite{Baratin:2011hp}.
As above, the covariance of the measure ensures the independence of the partition function on the normals,
which expresses the independence of the quantization on the gauge fixing.

\end{enumerate}

Several comments concerning this construction are in order:
\begin{itemize}
\item
These rules to construct the partition function
have been derived by discretizing the canonical path integral of the original theory and
by splitting it into contributions associated with different vertices \cite{Alexandrov:2008da}.
Therefore, the consistency with the canonical quantization is in a sense built-in to this approach.
In particular, we will see below how the quantum amplitudes introduced above allow to recover various elements of
loop quantum gravity.

\item
The key element of this construction is the formula for the vertex amplitude \eqref{amplit_simplex}.
This is a straightforward generalization of the usual prescription for the evaluation
of the vertex in the spin foam models considered in the literature. The only difference
is that the measure in \eqref{amplit_simplex} is supposed to be non-trivial and, in particular,
to include the secondary second class constraints. However, this difference has drastic consequences.
Without including the constraints, the integral defining the vertex is equivalent to the evaluation
of a simplex boundary state on a flat connection. This recipe has its origin in the topological BF theory
which describes the dynamics of flat connections. On the other hand, in the presence of the secondary constraints
in the measure, such correspondence does not work anymore and the dynamics becomes more complicated.

\item
While $A_v(\lambda_f,k_{e f},\inter_e)$ encodes the dynamics, the vertex amplitude in the ``connection" representation
$A_v[g_f,x_e]$ describes the kinematical Hilbert space of the theory. It shows that the kinematical states
always appear as linear combinations of projected spin networks \cite{Alexandrov:2007pq,Alexandrov:2008da},
which is indeed the case for all spin foam models of four-dimensional gravity considered in the literature.
But not all linear combinations are physically relevant. The space of all projected spin networks is too huge and only
the states given by \eqref{reprampl} contribute to the path integral.

\item
The non-commutative star-product is introduced in the final formula \eqref{fullpf-noncom} in order
to combine different exponential factors into the exponential of the discrete BF action.
This construction is similar to the one introduced recently in the study of a non-commutative flux
representation \cite{Baratin:2010wi,Baratin:2010nn,Baratin:2011hp}.

\item
It should be emphasized that at none of the steps one integrates over the normal vectors.
This is consistent with their interpretation as gauge fixing parameters. Moreover,
they completely drop out of the partition function, so that in the Lorentzian case an integral over
these normals would produce an overall infinite factor.

\item
An important feature of the proposed quantization is the difference in the roles of the primary and secondary constraints.
Whereas the latter enter the definition of the vertex amplitude and affect the dynamics,
the former become relevant only when one glues different simplices together, i.e. at the very last stage.
This is in a drastic contrast with the usual spin foam approach where the primary constraints play
the central role and the secondary constraints have been ignored at all so far.

\end{itemize}

\subsection{It works!}
\label{subsec-correct}

Let us now apply the quantization procedure of the previous subsection to our model
which, after partial gauge fixing, is given by the action \eqref{actiongauge-im}.
To make it more understandable, we will follow the procedure step by step.

\begin{enumerate}

\item
The second class constraints in our case are given by $\Phi^a_I$
and $\Psi_a^I$ \eqref{final-constr}. They have been already discretized in section \ref{subsec-discr}
(see \eqref{chiral-constr}), so that in the chiral notations we have
\be
\phi_{\rm discr}(B,x)=B_f^- -\xb_f B_f^+ \xb_f^{-1},
\qquad
\psi_{\rm discr}(g,x)=(g_{ve}^-)^{-1}\xb_v g_{ve}^+ \xb_e^{-1}.
\label{chiral-constr-discr}
\ee
It is useful to note that, besides the second class constraints on the canonical variables,
the gauge fixed action leads to conditions on the Lagrange multipliers.
These conditions, provided by \eqref{B0-x} and \eqref{condom0},
are naturally combined with the second class constraints,
which makes possible to extend the latter to spacetime covariant conditions.

\item
The measures \eqref{measur} take the form
\be
\begin{split}
\CD^{(x)} [B_f]&\,=\delta\(B_f^- -\xb_f B_f^+ \xb_f^{-1}\)\de B_f,
\\
\CD^{(x)} [g_{ve}]&\,=\delta\((g_{ve}^-)^{-1}\xb_v g_{ve}^+ \xb_e^{-1}\)\de g_{ve},
\end{split}
\label{measur-our}
\ee
where we took into account that the factor $\Delta(B,x)$ is trivial due to the last commutation relation
in \eqref{commPsi}. Note also that the measure over holonomies remains independent of the bivectors.

\item
The vertex amplitude given by the general formula \eqref{amplit_simplex} is evaluated in exactly
the same way as in section \ref{subsec-usual}.
Similarly to \eqref{gamma-amplit_simplex-model}, one finds
\be
A_v(\lambda_f,k_{e f},\inter_e)= \{15j\}\prod_{f\supset v} d_{k_f}^{-1} \delta_{k_{u(f)f}k_{d(f)f}}.
\label{amplit_simplex-model}
\ee
Thus, up to a normalization factor, we again reproduce the correct vertex amplitude of the $\SU(2)$
Crane-Yetter model \eqref{pfCY}.

\item
The rest of the construction will restore the correct face and edge amplitudes. But not only that.
It will also teach us several important lessons.
In particular, let us evaluate the vertex amplitude in the ``connection" representation \eqref{reprampl}.
Since the vertex \eqref{amplit_simplex-model} does not depend on $\lambda_f$, the sum over
this label can be done explicitly. Indeed, it is easy to prove the following identity
\be
\sum_{j^+,j^-}d_{j^+} d_{j^-}
\mathop{\sum}\limits_{m,m',n,n'}\ClG{j^+}{j^-}{j^{\mathstrut}}{m^{\mathstrut}}{m'}{\ell_1}\,
\Db^{(j^+)}_{mn^{\mathstrut}}(g^+)\Db^{(j^-)}_{m'n'}(g^-) \,
\overline{\ClG{j^+}{j^-}{j}{n^{\mathstrut}}{n'}{\ell_2}}
=\delta\( g^- (g^+)^{-1}\) \Db^{(j)}_{\ell_1 \ell_2}(g^+),
\label{property-state}
\ee
due to which one finds
\be
A_v[\gf_f,x_t]=\sum_{k_f,\inter_e} \{15j\}\[\prod_{f\supset v} d_{k_f}\,\delta\(\gf_f^-(x) (\gf^+_{f}(x))^{-1}\)\]
\bar\CS_{(\Gamma_v,k_f,\inter_e)}[\gf_f^+(x)],
\label{modelampl}
\ee
where we introduced $\gf_f(x)=g_{x_{u(f)}}^{-1} \gf_f g_{x_{d(f)}}^{\mathstrut}$
and $\CS_{(\Gamma_v,k_f,\inter_e) }$ is the usual $\SU(2)$ spin network associated with the boundary graph
$\Gamma_v$. This result has two remarkable features.
First, the delta function on the r.h.s. imposes the same condition as the second
class constraint \eqref{chiral-constr-discr} on the holonomy $g_{u(f)d(f)}$.
Second, the amplitude \eqref{modelampl} is represented as a sum over $\SU(2)$ spin networks,
which shows that the latter span the kinematical Hilbert space of our theory in full agreement with
the SU(2) Crane-Yetter model.
From this one concludes that the sum over the auxiliary representation labels, which the vertex amplitude does not depend on,
has two important effects: on the one hand, it restores the secondary second class constraints
on the arguments of the boundary states and, on the other hand,
it provides the reduction from the space of all projected spin networks
to the kinematical Hilbert space of the theory.
We emphasize that this reduction does not involve the primary simplicity constraints at all
and is achieved due to the secondary constraints encoded in the form of the vertex amplitude!

\item
The next step is to substitute \eqref{modelampl} into \eqref{amplitgen}. This gives
\be
A_v[\hat B_f,x_f]=\sum_{k_f,\inter_e} \{15j\}\prod_{f\supset v}
\[ d_{k_f}\int_{\SU(2)} \de h_{f}\,
e^{\I\tr\(b^{(\gamma)}_f h_{f}\)}\]
\bar\CS_{(\Gamma_v,k_f,\inter_e)}[h_{f}],
\label{amplitgen-our}
\ee
where we used the chiral decomposition of the $\so(4)$ trace \eqref{traces},
the gauge invariance of spin networks, and denoted
\be
\su(2)\,\ni\, b^{(\gamma)}_f=\hf\,(1+\gamma^{-1})B_f^+ +\hf\,(1-\gamma^{-1})\xb_f^{-1}B_f^-\xb_f  .
\ee

\item
Finally, we glue the amplitudes \eqref{amplitgen-our} together
by means of the formula \eqref{fullpf-noncom}.
To distinguish the representations and the group elements associated to different vertices,
we put the index $v$ on them, i.e. in \eqref{amplitgen-our} one should make the replacements
$(k_f,\inter_e)\mapsto (k_{vf},\inter_{ve})$ and $h_f\mapsto h_{vf}$. Then the non-commutative star-product
ensures that each face comes with the factor $\exp\left\{\I\tr(b^{(\gamma)}_f H_f)\right\}$
where $H_f=\prod_{v\subset f}h_{vf}$.
Since the geometric meaning of $h_{vf}$ is the positive chiral part of
the curvature around the wedge $(vf)$ (see Fig. \ref{fig-wedge}),
the group element $H_f$ gives (the positive chiral part of) the full curvature around the face $f$.
On the other hand, due to the primary simplicity constraints entering the measure on the bivectors,
one has $b^{(\gamma)}_f=B_f^+$ so that
the integrals in \eqref{fullpf-noncom} generate $\delta(H_f)$
imposing the flatness condition.
Expanding the $\delta$-function in the sum over representations, it is easy to see that
the full partition function is given by
\be
\CZ=\sum_{j_f}\sum_{k_{v f}, \inter_{ve}}\prod_{(v,f)} \[d_{k_{v f}}\int_{\SU(2)}\de h_{v f}\]
\prod_f \[d_{j_f}\chi_{j_f}\(\prod_{v\subset f}h_{v f} \)\]
\prod_v \[ \{15j\}\bar\CS_{(\Gamma_v,k_{vf},\inter_e)}[h_{v f}]\],
\ee
where $\chi_j$ is the $\SU(2)$ character of representation $j$.
It is immediate to check that, doing the remaining integration and contracting all indices,
one reproduces the $\SU(2)$ Crane-Yetter state sum \eqref{pfCY} with the same face, edge and vertex amplitudes.

\end{enumerate}

Thus, we conclude that the quantization rules given above lead to the correct spin foam quantization of
the degenerate sector of Plebanski theory and therefore provide the correct implementation of all the constraints.

\subsection{Why does it work?}
\label{subsec-expl}

In the previous subsection we went through a long way to get the partition function
of the constrained theory. In fact, in the particular case of our model, there is a shorter
way to arrive at the same result, which partially explains the origin of the proposed quantization rules.
It relies on the observation that the vertex amplitude in the ``connection" representation $A_v[\gf_f,x_e]$
has a much simpler expression. Indeed, in \cite{Alexandrov:2008da} it has been proven that
its spin foam like representation \eqref{reprampl} follows from
\be
A_v[\gf_f,x_e]=\int\prod_{e\supset v} \CD^{(x;B)} [\gl_{ve}]
\prod_{f\supset v}\delta\( \gl_{v u(f)}^{\mathstrut}\gf_f\gl_{vd(f)}^{-1}\).
\label{simbs_init}
\ee
As a result, one immediately obtains
\be
A_v[\hat B_f,x_f]=\int \prod_{e\supset v} \CD^{(x;B)} [\gl_{ve}]
\prod_{f\supset v}
\exp\(\I \Tr\[(1+\gamma^{-1} \star)B_f\cdot G_{vf}\]\),
\label{ampB}
\ee
where $G_{vf}=g_{fu(f)}\gl_{u(f)v}\gl_{vd(f)}g_{d(f)f}$ is the curvature around the wedge $(vf)$.
Then the partition function \eqref{fullpf-noncom} produces the exponential of the unconstrained $\Sp(4)$ BF action
integrated with the measure \eqref{measur} involving the primary and secondary second class constraints.
This is nothing else but a discretization of the standard canonical path integral for the constrained theory.
In our case the discrete constraints \eqref{chiral-constr-discr} can be solved explicitly,
which simply reduces the path integral to the $\SU(2)$ sector, and therefore the coincidence with the reduced phase space
quantization \eqref{pfCY} is guaranteed.

One can ask: why did we do all the above complicated calculations if they are not required to get the final spin foam model?
The point is that the shorter way is available only if it is possible to explicitly find the reduced phase space at the discrete level.
Although this can be done for our simple model, this seems to be out of reach in more complicated situations such as
the gravitational sector of Plebanski theory.
On the other hand, the representation \eqref{reprampl} disentangles the vertex contributions and
the gluing of different vertices, and importantly this is done {\it before} implementing the constraints.
This can be viewed as a first step towards the spin foam representation of the partition function,
which should follow after integrating out the remaining geometric variables.

Furthermore, the derivation of the previous subsection clarified many subtle issues such as
the imposition of constraints, gauge invariance, the role of the Immirzi parameter, etc.
Some of them have been already discussed above, and we will summarize once more our main conclusions
and observations in the next section.

\section{Discussion}

In this paper we studied the classical and quantum descriptions of the degenerate sector
of Plebanski formulation of general relativity. We have shown that one of its subsectors,
analogous to the gravitational sector of Plebanski theory, provides an interesting and useful model to
test the ideas of the spin foam quantization. Classically, it represents a constrained four-dimensional $\Sp(4)$
BF theory which, upon elimination of the constraints, reduces to the SU(2) BF theory.
Since both these theories are of BF type, their spin foam quantization is well known and can be used to find
the correct way of implementing the constraints.
As a result, we formulated a general procedure to build the partition function
of any constrained theory of Plebanski type. In particular, we showed that, being applied to our model,
this procedure works perfectly, giving rise to the right kinematical Hilbert space and generating the right dynamics,
i.e. those which agree with the SU(2) Crane-Yetter model.

One of the main results of this analysis is the clarification of the role of the primary and secondary second class constraints
in the construction of the spin foam partition function.
As has been already argued before \cite{Alexandrov:2008da,Alexandrov:2010pg,Geiller:2011kr,Alexandrov:2011km},
the secondary constraints affect the measure for holonomy variables and determine the form of the vertex
amplitude.\footnote{A similar modification of the measure for holonomies has been found also
in recent group field theory constructions \cite{Baratin:2011hp}.}
On the other hand, the primary constraints enter only at the very last step of the construction
when the contributions of different simplices are glued together.
In fact, this is a very natural result. The primary constraints is a simple consequence of the choice of
our basic variables $B_f$, which live on the boundary of 4-simplices. At the same time,
the secondary constraints appear as a commutator of the primary ones with the Hamiltonian
and therefore contain information about the dynamics of the theory. Moreover, they constrain the variables $g_{ve}$
living ``inside" 4-simplices. Giving these observations, it should not be surprising that
the main quantity responsible for the dynamics in the spin foam approach is governed by the secondary constraints,
whereas the primary ones play only some minor role at the boundary.
Furthermore, even the kinematical Hilbert space of the constrained theory
turns out to be completely determined by the secondary constraints because
the kinematical boundary states appear as projected spin networks weighted by the vertex amplitudes
(see \eqref{reprampl}).

All these statements are in a drastic contrast with the usual constructions performed
in the four-dimensional spin foam models of general relativity where the principal role is given to the primary constraints,
whereas the secondary constraints are not considered at all.
In the quasiclassical limit these models do provide the right dynamics
since in this limit one sets on shell where the secondary constraints are effectively induced.
However, beyond the limit the primary constraints are not sufficient to suppress the quantum fluctuations of the degrees
of freedom constrained by the secondary ones. This is why the latter should be taken into account
and are crucial to get the right dynamics at quantum level. This is clearly demonstrated by our model
as well as by its three-dimensional analogue \cite{Geiller:2011kr}.

It is worth also to note that our derivation confirmed once more that the closure constraint
of Regge calculus should not be imposed in spin foam models. This constraint requires
the invariance of intertwiners which is achieved by integration over the normal vectors $x_e$ associated to tetrahedra
of the simplicial decomposition. However, our model, in agreement with previous claims
\cite{Alexandrov:2007pq,Alexandrov:2008da,Rovelli:2010ed}, clearly shows that such integration would be inconsistent
and the normal vectors should be kept fixed, which is nothing else but the usual gauge fixing of the boost
gauge freedom in the gravity path integral.

Another interesting point is that our quantization has been done in the presence of the Immirzi parameter.
Nevertheless, it did not have any effect on the constrained theory both at classical and quantum level.
After imposition of the constraints,
it completely drops out of the partition function. This should be compared with the claims that
its appearance in loop quantum gravity is a consequence of an unfortunate choice of variables
(the Ashtekar-Barbero connection which is not a pull-back of a spacetime connection)
and with a right choice it disappears from physical results
\cite{Alexandrov:2001pa,Alexandrov:2001wt,Alexandrov:2010un}.

Finally, let us comment on the extension of our construction to the physical case of the gravitational sector
of Plebanski theory.
The main difference distinguishing it from our model is the form of the constraints which cannot be written anymore
in the simple form \eqref{final-constr} or \eqref{chiral-constr-discr} after discretization.
In particular, the secondary constraints become explicitly dependent on the $B$-field
\cite{Buffenoir:2004vx,Alexandrov:2006wt}. Although the construction of section \ref{subsec_lessons}
is still well defined in the presence of such dependence, it gives rise to many complications.
The most important one is that the quantity \eqref{amplit_simplex} starts to depend on the bivectors $B_f$
and its interpretation as a vertex amplitude is not viable anymore.
It is not clear whether this is a serious problem or just a minor obstacle.
In principle, the $B$-dependence can make impossible to integrate out the bivectors because
the resulting integrals are not of $BF$ type anymore. However, given that in our procedure this integration appears
only as a way to glue the simplex contributions, one may still hope that it will be possible to factorize
the $B$-dependence and to extract the spin foam vertex.
In any case, this issue certainly deserves a further study.

\section*{Acknowledgements}

The author is grateful to Aristide Baratin, Winston Fairbairn, Marc Geiller, Karim Noui,
Daniele Oriti and Simone Speziale for useful discussions and correspondence, as well as
to two anonymous referees for various suggestions which helped to improve the presentation.
This research is supported by contract ANR-09-BLAN-0041.

\appendix

\section{Conventions}
\label{ap-conven}

\subsection{ {$\so(4)$} algebra and the chiral decomposition}
\label{ap-algernra}

Our conventions for indices are the following: $\mu,\nu,\dots$ denote spacetime indices,
$a,b,\dots$ are spatial indices, $I,J,\dots$ are indices in the tangent space which carries
the fundamental representation of $\so(4)$, and $i,j,\dots$
refer to the $\su(2)$ subalgebra.
$\eps^{IJKL}$ and $\eps^{ijk}$ are the invariant fully anti-symmetric tensors in four and three dimensions,
respectively, normalized by $\eps^{0123}=\eps^{123}=1$.
Besides, we use $(\cdot\cdot)$ and $[\cdot\cdot]$ to denote
symmetrization and anti-symmetrization, respectively, with weight 1/2.
Since all tangent space indices are raised and lowered with the metric $\eta_{IJ}=\diag(1,1,1,1)$,
we do not follow the rule, that the contracted indices should be in opposite positions, very strictly.

The canonical basis of $\so(4)$ is composed of the rotation and boost generators, $L_i$ and $K_i$, which have
the following commutation relations
\be
[L_i,L_j]={\eps_{ij}}^k L_k,
\qquad
[K_i,K_j]={\eps_{ij}}^k L_k,
\qquad
[K_i,L_j]={\eps_{ij}}^k K_k.
\label{comm-canon}
\ee
The rotation generators form the canonically embedded $\su(2)$ subalgebra.
On the other hand, given a four-dimensional normal vector $x^I$, one can introduce a boosted subalgebra $\su(2)_x$.
It is formed by the generators which leave the vector invariant.
If one introduces the covariant notation for the $\so(4)$ generators $T^{ij}=\hf\,\eps^{ijk}L_k$, $T^{0i}=\hf\,K^i$,
then
\be
\begin{split}
I^{IJ}_{KL}(x)=&\,\hf\, \eps^{IJMP}\eps_{KLNP}x^M x_N=\delta^{IJ}_{KL}-2 x^{[J}\delta^{I]}_{[K} x_{L]},
\\
J^{IJ}_{KL}(x)=&\,2 x^{[J}\delta^{I]}_{[K} x_{L]}.
\end{split}
\label{proj}
\ee
are the projectors on $\su(2)_x$ and its orthogonal completion, respectively,
i.e. acting on the generators in the vector representation, they satisfy $I^{IJ}_{KL}(x) T^{KL}\cdot x^N=0$.

The antisymmetric bivectors $B^{IJ}$ form the adjoint representation of the $\so(4)$ algebra.
On this representation we define the action of the Hodge operator
as $(\star B)^{IJ}=\hf\,{\eps^{IJ}}_{KL}B^{KL}$.
Since the Hodge operator squares to one, $\star^2={\rm id}$, it splits the space of
bivectors into the direct sum of two eigenspaces with eigenvalues $\pm 1$,
\be
B=B^{(+)}_iT^{(+)}_i +B^{(-)}_iT^{(-)}_i,
\qquad
\star T^{(\pm)}=\pm T^{(\pm)}.
\label{chiraldecomp}
\ee
This corresponds to the chiral decomposition $\Sp(4)=\SU(2)\times \SU(2)$, and the generators of the Lie algebras
of the two chiral subgroups are given by $T^{(\pm)}_i=\hf\,(L_i\pm K_i)$ with
\be
[T^{(\pm)}_i,T^{(\pm)}_j]={\eps_{ij}}^k T^{(\pm)}_k,
\qquad
[T^{(+)}_i,T^{(-)}_j]=0.
\ee

In the adjoint representation we normalize the trace such that
\be
\Tr(T^{IJ} T_{KL})=\delta^{IJ}_{KL}
\ \Longrightarrow\
\Tr(AB)=A^{IJ} B_{IJ},
\ee
where $\delta^{IJ}_{KL}=\delta^{[I}_{K}\delta^{J]}_L$. Then the chiral decomposition \eqref{chiraldecomp} implies that
\be
\Tr(T^{(\epsilon)}_iT^{(\epsilon')}_j)=\delta_{\epsilon\epsilon'}\delta_{ij},
\ \Longrightarrow\
\Tr(AB)=A^{(+)}_i B^{(+)}_i+A^{(-)}_i B^{(-)}_i.
\ee
On the other hand, it is useful to remember that the $\su(2)$ generators $L_i$ have a different normalization,
namely, $\tr(L_i L_j)=2\delta_{ij}$. Due to this, in terms of the $\su(2)$ elements $B^{(\pm)}=B^{(\pm)}_i L^i$,
the $\so(4)$ trace reads
\be
\Tr(AB)=\hf\,\tr(A^{(+)}B^{(+)})+\hf\,\tr(A^{(-)}B^{(-)}).
\label{traces}
\ee

\subsection{Clebsch-Gordan coefficients}
\label{ap-ClG}

Our conventions for $\SU(2)$ invariant intertwiners follow \cite{Freidel:2007py}. A generic invariant
intertwiner is denoted by $\inter$ and is supposed to be normalized as
\be
\sum_{m_1\cdots m_L}\inter_{m_1\cdots m_L}\overline{\inter_{m_1\cdots m_L}}=1.
\label{normali}
\ee
In the particular case of three coupled representations, the matrix elements of the intertwiners
are given by the Clebsch-Gordan coefficients $\ClG{j_1}{j_2}{j_3}{m_1}{m_2}{m_3}$.
It is convenient also to define the two invariant maps based on these intertwiners
\be
\ClG{j_1}{j_2}{j_3}{}{}{}\ :\ \CH^{(j_1)}\otimes\CH^{(j_2)}\rightarrow \CH^{(j_3)}
\quad \text{and}\quad
\overline{\ClG{j_1}{j_2}{j_3}{}{}{}}\ :\ \CH^{(j_3)} \rightarrow\CH^{(j_1)}\otimes\CH^{(j_2)}.
\ee
Then the following properties are satisfied
\be
\ClG{j_1}{j_2}{j}{}{}{}\overline{\ClG{j_1}{j_2}{j'}{}{}{}}=d_j^{-1}\delta_{jj'}\unit_j,
\qquad
\sum_{j=|j_1-j_2|}^{j_1+j_2}d_j \overline{\ClG{j_1}{j_2}{j}{}{}{}}\ClG{j_1}{j_2}{j}{}{}{}
=\unit_{j_1}\otimes\unit_{j_2},
\ee
where $d_j=2j+1$ is the dimension of the $\SU(2)$ representation.
In particular, the first property ensures that one gets the right normalization \eqref{normali}.
Finally, we fix the normalization of Wigner matrices by requiring
\be
\int_{\SU(2)} \de h\,\Db^{(j)}_{mn}(h)\overline{\Db^{(j')}_{m'n'}(h)}=d_j^{-1}\delta_{jj'}\delta_{mm'}\delta_{nn'}.
\ee
With these normalizations, the matrix elements are recoupled as follows
\be
\Db^{(j_1)}_{m_1n_1}(h)\Db^{(j_2)}_{m_2n_2}(h)=\sum_{j=|j_1-j_2|}^{j_1+j_2}\sum_{m,n}
d_j \overline{\ClG{j_1}{j_2}{j}{m_1}{m_2}{m}}\ClG{j_1}{j_2}{j}{n_1}{n_2}{n}\Db^{(j)}_{mn}(h).
\ee
These properties are sufficient to prove \eqref{property} and \eqref{property-state}.

\section{Canonical analysis of the degenerate Plebanski sector}
\label{ap-canan}

The canonical analysis of the gravitational sector of $\Sp(4)$ Plebanski theory
has been carried out for the first time in \cite{Buffenoir:2004vx},
and further elaborated in \cite{Alexandrov:2006wt,Alexandrov:2008fs}.
The degenerate sector described by the action \eqref{plebanskiaction2} can be analyzed along
a similar way. Here we follow the original method of \cite{Buffenoir:2004vx}.

One starts as usual from the 3+1 decomposition. One immediately observes that
$\omega_{a}^{IJ}$ and $\tP^a_{IJ}=\eps^{abc}B_{bc}^{IJ}$ appear as conjugate variables, whereas
the action does not contain time derivatives of $\lambda^{\mu\nu\rho\sigma}$, $\omega_{0}^{IJ}$
and $B_{0a}^{IJ}$ which therefore are expected to play the role of Lagrange multipliers.
However, the latter variables appear quadratically in the term generating the simplicity constraints
which leads to certain complications in considering them directly as Lagrange multipliers.
To avoid these complications. one adds new non-dynamical variables $\mu_a^{IJ}$ and $\pi^a_{IJ}$ which
enforce the vanishing of the momenta conjugate to $B_{0a}^{IJ}$ by means of the following additional term
\be
\int_\CM\de^4\xc\Big(\Tr(\pi^a\partial_0 B_{0a})-\Tr(\mu_a\pi^a)\Big).
\ee
Then the pase space is spanned by $\omega_a^{IJ}$, $B_{\mu\nu}^{IJ}$ and $\pi_a^{IJ}$
with the symplectic structure given by
\be
\{\omega_a^{IJ}(\xc), \tP^b_{KL}(\yc)\}=\delta_a^b \delta^{IJ}_{KL}\delta(\xc,\yc),
\qquad
\{B_{0a}^{IJ}(\xc), \pi^b_{KL}(\yc)\}=\delta_a^b \delta^{IJ}_{KL}\delta(\xc,\yc),
\ee
and the total action leads to the following primary constraints
\begin{subequations}
\beq
\pi^a_{IJ} &\approx& 0
\\
\CG_{IJ}&=& D_a \tP^a_{IJ}\approx 0,
\\
\Phi(B,B)_{ab}&=&\hf\,\eps_{IJKL} B_{0a}^{IJ}B_{0b}^{KL}\approx 0,
\label{cBB}
\\
\Phi(\tP,B)^a_{~b}&=&\hf\, {\eps^{IJ}}_{KL}\tP^a_{IJ} B_{0b}^{KL}\approx 0,
\label{cPB}
\\
\Phi(\tP,\tP)^{ab}&=&\hf\,\eps^{IJKL}\tP^a_{IJ}\tP^b_{KL}\approx 0.
\label{cPP}
\eeq
\label{primarycon}
\end{subequations}
The last three are nothing else but the various components of the simplicity constraints \eqref{simplicity constraint}.
But as is shown in appendix \ref{ap-constr}, the six constraints $\Phi(B,B)_{ab}$ are in fact linear combinations
of $\Phi(\tP,B)^a_{~b}$ and $\Phi(\tP,\tP)^{ab}$, and do not require a special attention.
The Hamiltonian can be written in the following form
\be
\begin{split}
-H =&\, \eps^{abc}B_{oa}^{IJ} F_{bc}^{IJ}+\omega_0^{IJ} \CG_{IJ}
+\lambda^{0a0b}\Phi(B,B)_{ab}+\hf\,\lambda^{0abc}\eps_{bcd}\Phi(\tP,B)^d_{~a}
\\
&\,
+\frac{1}{16}\,\lambda^{abcd}\eps_{abf}\eps_{cdg}\Phi(\tP,\tP)^{fg}
-\mu_a^{IJ}\pi^a_{IJ}.
\end{split}
\label{Hamilt}
\ee

The next step is to study the conditions imposed by the conservation of the primary constraints.
Let us compute their time derivatives by commuting them with the Hamiltonian \eqref{Hamilt}.
First, one finds
\be
\dot \pi^a_{IJ}=\eps^{abc}F_{bc}^{IJ}+2\lambda^{0a0b}(\star B)_{0b}^{IJ}
+\hf\,\lambda^{0abc}\eps_{bcd}(\star\tP)^d_{IJ}\approx 0.
\label{pres-pi}
\ee
These 18 equations split into two sets.
By contracting them with $\tP^b_{IJ}$, the last terms produce the simplicity constraints
so that one remains with 9 conditions
\be
\CC^{ab}=\eps^{acd}F_{cd}^{IJ}\tP^b_{IJ}\approx 0,
\label{secconC}
\ee
which should be interpreted as secondary constraints.
The remaining 9 equations following from \eqref{pres-pi} can be written as
\be
\eps^{bcd}\Tr(\tP^a\star F_{cd})+
2\lambda^{0b0c}\Tr(\tP^a B_{0c})
+\hf\,\lambda^{0bcd}\eps_{cdg}\Tr(\tP^a\tP^g)=0
\ee
and fix the Lagrange multipliers $\lambda^{0bcd}$.

The conservation of the Gauss constraint $\CG_{IJ}$ does not generate new conditions
provided it is shifted as follows
\be
\CG'_{IJ}=\CG_{IJ}+[B_{0a},\pi^a]_{IJ}.
\ee
The shift ensures that $\CG'_{IJ}$ is a generator of $\Sp(4)$ gauge transformations
and since the Hamiltonian is gauge invariant, $\CG'_{IJ}$ is preserved under evolution.

Next it is convenient to consider $\Phi(\tP,\tP)^{ab}$. Its commutator with the Hamiltonian
generates 6 new conditions
\be
\Psi^{ab}=\eps^{cd(a}\Tr(B_{0d}\star D_c\tP^{b)})\approx 0,
\label{seccl-cosntr}
\ee
which give rise to additional secondary constraints. Having obtained these constraints,
we can now prove a very useful Lemma which facilitates a lot the following analysis.

{\bf Lemma:} {\it Let $b_{\mu\nu}^I$ be defined by the solution of the simplicity constraints in any of the two sectors
(see \eqref{twosectors}).
Assume that i) $\tp^a_I=\eps^{abc}b_{bc}^I$ is invertible in the sense that there exists $\pt_a^I$ such that
\be
\pt_a^I\tp^b_I=\delta_a^b,
\qquad
\pt_a^I\tp^a_J=\delta^I_J-x^Ix_J;
\ee
ii) the matrix
\be
\CQ^{ab,cd}={\eps^{IJK}}_L x_I \tp^{(a}_J\eps^{b)g(c}\tp^{d)}_Kb_{0g}^L
\ee
in invertible. Then $D_a x^I$ can be expressed as a linear combination of $\Psi^{ab}$, $\tG^a\equiv\Tr(\tP^a\star\CG)$
and the simplicity constraints, and therefore it weakly vanishes.}

{\bf Proof:}
First, it is easy to see that in both solution sectors \eqref{twosectors}, i.e. on the surface of the simplicity constraints,
one has
\be
\begin{split}
\tG^a &\, =-\hf\, \eps^{IJKL} x_I \tp^a_J\tp^b_K D_b x_L,
\\
\Psi^{ab} &\, =-\hf\,\eps^{cd(a} {\eps_{IJ}}^{KL} x^I b_{0d}^J\tp^{b)}_K D_c x_L.
\end{split}
\ee
Then one can check that
\be
\CQ^{ab,cd}\pt_{(c}^I D_{d)} x_I=2\Psi^{ab}-\tG^{(a}\eps^{b)cd} b_{0c}^I\pt_{d,I}.
\ee
Since the matrix $\CQ^{ab,cd}$ is assumed to be invertible, this implies that $\pt_{(a}^I D_{b)} x_I$ is weakly vanishing.
Finally, one verifies that
\be
D_a x_I=\tp^b_I\(\pt_{(a}^J D_{b)} x_J\)-{\eps_{IJKL}}x^J\pt_a^K\pt_b^L\tG^b\approx 0.
\ee
$\Box$

We still need to analyze the stability of two simplicity constraints, \eqref{cBB} and \eqref{cPB}.
However, since the former is expressible through the latter and \eqref{cPP}, only $\Phi(\tP,B)^a_{~b}$
remains to be considered. Its time derivative leads to the following condition
\be
\dot\Phi(\tP,B)^a_{~b}=
-\Tr([\omega_{0}, B_{0b}]\star \tP^a) +\Tr(\mu_b \star\tP^a)
\approx 0,
\label{cond-mu}
\ee
where we neglected the term $2\eps^{acd}\Tr(B_{0b}\star D_c B_{od})$ because, on the surface of the simplicity constraints,
it is proportional to $D_c x^I$ and vanishes weakly by the above Lemma.
The resulting equation fixes 9 of the 18 components of the Lagrange multiplier $\mu_a^{IJ}$
and does not lead to new constraints.

This completes the analysis of the primary constraints.
But the appearance of the secondary constraints $\CC^{ab}$ and $\Psi^{ab}$ requires
to repeat the stabilization procedure.
However, at this step it works rather differently for the deg-gravitational and deg-topological sectors.
Due to this reason, we consider them separately.

\subsection{Degenerate gravitational sector}

First, let us consider the conservation of the secondary constraints $\Psi^{ab}$.
This gives the following equation
\be
\begin{split}
\dot\Psi^{ab}=&\,
2\eps^{cd(a}\eps^{b)fg}\Tr(B_{0d}\star D_cD_f B_{0g})
-\eps^{cd(a}\Tr([\omega_0,B_{0d}]\star D_c\tP^{b)})
+ \eps^{cd(a}\Tr(\mu_{d}\star D_c\tP^{b)})
\\
&\,
+\lambda^{0cd(a}\Tr(\tP^{b)}[B_{0c},B_{0d}])
+\frac{1}{8}\,\lambda^{fgpq}\eps_{cfg}\eps_{rpq}\eps^{cd(a}\Tr(\tP^{b)}[\tP^r,B_{0d}])
\approx 0.
\end{split}
\label{preservPsi}
\ee
However, it is easy to see that the first term weakly vanishes due to the Lemma, whereas
the next two terms produce the equation \eqref{cond-mu} plus contributions proportional to $D_c x^I$.
As a result, the stability condition reduces to the vanishing of the last two terms.
The crucial question for us is the form of the matrix in front of the Lagrange multiplier
$\lambda^{fgpq}$ in the last term, which arises from the commutator of $\Psi^{ab}$
with the primary constraints $\Phi(\tP,\tP)^{cd}$. In the deg-gravitational sector,
where the $B$-field is given by \eqref{gravsector}, it is found to be
\be
\{\Psi^{ab},\Phi(\tP,\tP)^{cd}\}=
4\tP^{(a}_{IJ}\,\eps^{b)g(c}\,\tP^{d)}_{IK}B_{0g}^{JK}\ \mathop{\approx}\limits^{\text{grav}}\
\CQ^{ab,cd}.
\label{commPhiPsi}
\ee
Since this is the same matrix which appears in the Lemma and generically it is invertible,
the condition \eqref{preservPsi} fixes the six Lagrange multipliers $\lambda^{fgpq}$. At the same time, it shows
that $\Phi(\tP,\tP)^{ab}$ and $\Psi^{ab}$ are mutually non-commuting.

Before we proceed further, we prove an additional useful result that the curvature of the spin connection
is weakly vanishing. Indeed, using \eqref{gravsector} and the inverse field $\pt_a^I$ introduced above,
one has
\be
\begin{split}
\eps^{abc}F_{bc}^{IJ}&\,=\eps^{abc}I^{IJ}_{KL}(x)F_{bc}^{KL}+\eps^{abc}J^{IJ}_{KL}(x)F_{bc}^{KL}
\\
&\, ={\eps^{IJ}}_{KL}x^K \pt_b^L\CC^{ab}+4x^{[J}\eps^{abc}D_bD_c x^{I]}\approx 0,
\end{split}
\label{Fzero}
\ee
where in the second term we represented the curvature as a commutator of two covariant derivatives.
Evaluating now the time derivative of the secondary constraint $\CC^{ab}$
\be
\dot\CC^{ab}=2\eps^{acd}\eps^{bgf}\Tr(B_{0g}D_fF_{cd})
-2\lambda^{0gac}\Tr(\tP^b\star D_c B_{og})
-\frac{1}{4}\, \lambda^{fgpq}\eps_{cfg}\eps_{rpq}\eps^{cda}\Tr(\tP^{b}\star D_d\tP^r)\approx 0,
\label{timeCC}
\ee
one immediately concludes that the stability condition is satisfied due to the above Lemma and
the vanishing of the curvature proven in \eqref{Fzero}.

As a result, the stabilization procedure finishes at this point and the list of all constraints
is given by $\pi^a_{IJ}$, $\CG_{IJ}$, $\Phi(\tP,B)^a_{~b}$, $\Phi(\tP,\tP)^{ab}$, $\Psi^{ab}$, and $\CC^{ab}$.
Note however that the last constraints are reducible. Namely, due to the Bianchi identity, they satisfy
\be
D_a( \CC^{ab}\pt_b^I)=\hf\,\eps^{IJKL}\eps^{acd} F_{cd}^{KL}D_a x_J,
\ee
where the r.h.s., as we know, vanishes on the surface of the other constraints.
Thus, only six components of $\CC^{ab}$ are independent.
To split the resulting constraints into first and second class, we introduce
\be
\piI^{ab}=\Tr(\pi^a\tP^b),
\qquad
\piII^{ab}=\Tr(\pi^a\star\tP^b).
\label{newpi}
\ee
Then, using the weak vanishing of the curvature and of the covariant derivative of the normal vector,
it is straightforward to verify that $\piI^{ab}$, $\CG'_{IJ}$ and $\CC^{ab}$
are first class, whereas $\piII^{ab}$, $\Phi(\tP,B)^a_{~b}$, $\Phi(\tP,\tP)^{ab}$ and $\Psi^{ab}$
are second class. This implies the following counting of degrees of freedom.
The original phase space is $4\times 18=72$ dimensional.
The second class constraints remove $9+9+6+6=30$ degrees of freedom, whereas the first class constraints
together with the corresponding gauge fixing conditions fix $2\times(9+6+(9-3))=42$ of them. This leaves us with a zero-dimensional
phase space confirming that we are describing a topological theory.

If one partially fixes the gauge, taking the normal vector $x^I$ to be a given function $x^I(\xc)$,
this gauge fixing condition can be combined with $\Phi(\tP,\tP)^{ab}$ to get the simplicity constraints
\eqref{Phi-x}. Moreover, the part of the Gauss constraint $\tG^a$ is conjugate to the gauge fixing condition
and therefore becomes second class. As shown by our Lemma, it can be combined with $\Psi^{ab}$
to get the secondary constraints \eqref{secondarygauge}. Besides, the constraints $\Phi(\tP,B)^a_{~b}$
become identical to the condition \eqref{B0-x} on the Lagrange multipliers, whereas $\CC^{ab}$ and
the remaining part of $\CG_{IJ}$ are equivalent to $\CC^a_I$ and $\hat \CG_I$ defined in \eqref{constrCC}
and \eqref{rem-constr}, respectively. After integrating out the auxiliary variables $\pi^a_{IJ}$,
one recovers the same phase space and the same constraint structure as in section \ref{subsec-canan}
where the canonical analysis was performed for the gauge fixed action \eqref{actiongauge}.

Furthermore, it is possible to show the following equality
\be
\begin{split}
&
{\bf \Delta}_{\rm D}^{1/2}\left|\det\{x,\tG\}\right|\,
\delta(\piI)\delta(\piII)\delta\(\Phi(\tP,\tP)\)\delta\(\Phi(\tP,B)\)\delta(\Psi)\delta(\tG)\delta(x-x(\xc))
\\
& \sim
\delta(\pi^a_{IJ})\delta(x_J B_{0a}^{IJ})\delta(x^J \tP^a_{IJ})\delta(D_a x^I),
\end{split}
\ee
where
\be
{\bf \Delta}_{\rm D}\sim (\det\tp)^6(\det Q)^2
\ee
is the determinant of the Dirac matrix of the commutators of the second class constraints,
the second factor is a part of the Faddeev-Popov determinant corresponding to the gauge fixing of
the normal $x^I$, and all equations are given up to numerical factors.
The meaning of this result is twofold: first, it demonstrates explicitly the recombination of
the constraints mentioned above and, second, it confirms the triviality of the factor $\Delta$ from \eqref{measur}
(see \eqref{measur-our}). Thus, all results derived from the gauge fixed action \eqref{actiongauge}
are in the full agreement with the complete canonical analysis of the initial gauge invariant theory.

\subsection{Degenerate topological sector}

Now we turn to the deg-topological sector described by the solution \eqref{tpsector}.
The conservation of the secondary constraints $\Psi^{ab}$ still generates the condition \eqref{preservPsi}
where the first line can be dropped.
However, in contrast to the previous case, the matrix appearing in the last term vanishes
\be
\{\Psi^{ab},\Phi(\tP,\tP)^{cd}\}=
\tP^{(a}_{IJ}\,\eps^{b)g(c}\,\tP^{d)}_{IK}B_{0g}^{JK}\ \mathop{\approx}\limits^{\text{top}}\
0
\label{commPhiPsi-zero}
\ee
so that $\Phi(\tP,\tP)^{ab}$ and $\Psi^{ab}$ are now mutually commuting. Moreover, the forth term also vanishes
in the deg-topological sector so that the stability of $\Psi^{ab}$ does not generate any new conditions.

The second crucial difference arising in this sector is that the secondary constraints $\CC^{ab}$ \eqref{secconC}
turn out to be linearly dependent with other constraints. Indeed, using \eqref{tpsector} and
the same trick as in \eqref{Fzero}, one obtains
\be
C^{ab}=-2\eps^{acd}\tp^b_I D_c D_d x^I.
\ee
Then the above Lemma ensures that this quantity can be expressed as a linear combination of the Gauss constraint,
simplicity and $\Psi^{ab}$. Due to this, $C^{ab}$ does not require a separate consideration and its stability follows
from the above analysis.

As a result, the independent set of constraints is provided by
$\pi^a_{IJ}$, $\CG_{IJ}$, $\Phi(\tP,B)^a_{~b}$, $\Phi(\tP,\tP)^{ab}$, and $\Psi^{ab}$.
To split them into first and second class, besides \eqref{newpi}, we define
\be
\hat\Psi^{ab}=\Psi^{ab}-\{\Psi^{ab},\piII^{cd}\}\(\Tr(\tP\tP)^{-1}\)_{df}\Phi(\tP,B)^f_{~c}.
\ee
Then one can check that $\piI^{ab}$, $\CG'_{IJ}$, $\Phi(\tP,\tP)^{ab}$ and $\hat\Psi^{ab}$
are first class, whereas $\piII^{ab}$ and $\Phi(\tP,B)^a_{~b}$ are mutually non-commuting and therefore
give second class constraints.
The counting of degrees of freedom works as above.
The second class constraints give $9+9=18$ conditions, whereas the first class constraints
and their gauge fixing conditions produce $2\times(9+6+6+6)=54$ more. Altogether they fix all degrees of freedom
of the original 72-dimensional phase space, also showing that this sector describes a topological theory.

\section{Constraints for constraints}
\label{ap-constr}

Let us consider the simplicity constraints split according to the 3+1 decomposition as in \eqref{primarycon}
and written in terms of the chiral variables. They take the following form
\be
\begin{split}
\Phi(B,B)_{ab}&\,=\Bp{a}{i}\Bp{b}{i}-\Bm{a}{i}\Bm{b}{i},
\\
\Phi(\tP,B)^a_{~b}&\,=\tPp{a}{i}\Bp{b}{i}-\tPm{a}{i}\Bm{b}{i},
\\
\Phi(\tP,\tP)^{ab}&\,=\tPp{a}{i}\tPp{b}{i}-\tPm{a}{i}\tPm{b}{i}.
\end{split}
\label{symcosntr}
\ee
Then it is straightforward to check that the following linear combination of constraints
\be
\begin{split}
\ccon_{ab}=&\,
\eps_{c_1c_2c_3}\eps_{d_1d_2d_3}\[  \frac12\(\frac13\,\tPp{c_1}{i_1}\tPm{d_1}{i_1}\Phi(B,B)_{ab}
-\Bp{a}{i_1}\tPm{d_1}{i_1}\Phi(\tP,B)^{c_1}_{~b}
\right.\right.
\\
&\,
\left.\left.
-\tPp{c_1}{i_1}\Bm{b}{i_1}\Phi(\tP,B)^{d_1}_{~a}
+\Bp{a}{i_1}\Bm{b}{i_1}\Phi(\tP,\tP)^{c_1d_1}\)
\tPp{c_2}{i_2}\tPp{c_3}{i_3}\tPm{d_2}{i_2}\tPm{d_3}{i_3}
\right.
\\
&\, \left.
-\Bp{a}{i_2}\tPp{c_2}{i_1}\tPp{c_3}{i_3}\Bm{b}{i_1}\tPm{d_2}{i_2}\tPm{d_3}{i_3}\Phi(\tP,\tP)^{c_1d_1}
\]
\end{split}
\label{conforcon}
\ee
identically vanishes. This can be done either expanding explicitly all sums over repeated indices
or assuming that $\tPpm{a}{i}$ are two invertible matrices which allows to write, for instance,
\be
\eps_{c_1c_2c_3}\tPp{c_2}{i_2}\tPp{c_3}{i_3}=\(\det\tPp{}{}\){\eps_{i_1}}^{i_2i_3}\(\tPp{-1}{}\)_{c_1}^{i_1}.
\ee
Using this property in each term in \eqref{conforcon}, the vanishing of $\ccon_{ab}$ follows trivially.

The nine quantities $\ccon_{ab}$ encode constraints for the simplicity constraints \eqref{simplicity constraint}.
However, they themselves are not all independent.
If one considers their antisymmetric part $\ccon_{[ab]}$, it is possible to show that the coefficients in front of
the constraints vanish on the constraint surface. Indeed, assuming again the invertibility of $\tPpm{a}{i}$, one finds
\be
6\ccon_{[ab]}=\( \Bp{[b}{i}\(\tPp{-1}{}\)_{c}^{i}-\Bm{[b}{i}\(\tPm{-1}{}\)_{c}^{i}\)
\(\Phi(\tP,B)^c_{~a]} - \Bm{a]}{j}\(\tPm{-1}{}\)_{d}^{j}\Phi(\tP,\tP)^{cd}\),
\ee
where the prefactor can be rewritten as a linear combination of the original constraints \eqref{symcosntr}
\be
\begin{split}
&
\Bp{b}{i}\(\tPp{-1}{}\)_{c}^{i}-\Bm{b}{i}\(\tPm{-1}{}\)_{c}^{i}=
\\
&\qquad\qquad
2\(\tr(\tPp{}{}\tPp{}{})^{-1}\)_{cd}\(\Phi(\tP,B)^d_{~b}-\Bm{b}{i}\(\tPm{-1}{}\)_{g}^{i}\Phi(\tP,\tP)^{gd}\).
\end{split}
\ee
Thus, only the symmetric part of \eqref{conforcon} generates constraints for constraints
and allows to express $\Phi(B,B)_{ab}$ in terms of other 15 simplicity constraints.

One can also notice that the vanishing of \eqref{conforcon} continues to hold even if one replaces there
one or two of the fields $B_{0a}$ by $\tP^a$. For example, replacing only one field, one
obtains the following combination of constraints
\be
\begin{split}
\hat \ccon^a_{~b}=&\,
\eps_{c_1c_2c_3}\eps_{d_1d_2d_3}\[  \frac12\(\frac13\,M^{c_1 d_1}\Phi(\tP,B)^a_{~b}
-M^{ad_1}\Phi(\tP,B)^{c_1}_{~b}-\tPp{c_1}{i}\Bm{b}{i}\Phi(\tP,\tP)^{ad_1}
\right.\right.
\\
&\,
\left.\left.
+\tPp{a}{i}\Bm{b}{i}\Phi(\tP,\tP)^{c_1d_1}\)
M^{c_2 d_2}M^{c_3d_3}
-\tPp{c_2}{i}\Bm{b}{i} M^{ad_2}M^{c_3d_3}\Phi(\tP,\tP)^{c_1d_1}
\],
\end{split}
\label{conforcon-no}
\ee
where we denoted $M^{ab}=\hf\,\tr(\tPp{}{}\tPm{}{})^{ab}$. This combination is also vanishing, so that one could think
that these are additional constraints for constraints. However, in contrast to \eqref{conforcon},
$\hat \ccon^a_{~b}$ vanish for any $\Phi(\tP,B)^a_{~b}$ and symmetric $\Phi(\tP,\tP)^{ab}$, not necessarily
of the form \eqref{symcosntr}. Again this can be easily proven by assuming the invertibility of the matrix $M^{ab}$.
Due to this, they do not reduce the number of independent constraints which remains to be 15.

\providecommand{\href}[2]{#2}\begingroup\raggedright\endgroup


\begin{thebibliography}{10}

\bibitem{Baez:1997zt}
J.~C. Baez, ``{Spin foam models},''
  \href{http://dx.doi.org/10.1088/0264-9381/15/7/004}{{\em Class. Quant. Grav.}
  {\bf 15} (1998)  1827--1858},
\href{http://arxiv.org/abs/gr-qc/9709052}{{\tt arXiv:gr-qc/9709052}}.

\bibitem{Perez:2003vx}
A.~Perez, ``{Spin foam models for quantum gravity},'' {\em Class. Quant. Grav.}
  {\bf 20} (2003)  R43,
\href{http://arxiv.org/abs/gr-qc/0301113}{{\tt arXiv:gr-qc/0301113}}.

\bibitem{Ponzano:1968dq}
G.~Ponzano and T.~Regge. In ``Spectroscopy and group theoretical methods in
  physics'', ed. by F. Block (North Holland, 1968).

\bibitem{Turaev:1992hq}
V.~Turaev and O.~Viro, ``{State sum invariants of 3 manifolds and quantum 6j
  symbols},'' \href{http://dx.doi.org/10.1016/0040-9383(92)90015-A}{{\em
  Topology} {\bf 31} (1992)  865--902}.

\bibitem{Witten:1988hf}
E.~Witten, ``{Quantum Field Theory and the Jones Polynomial},''
  \href{http://dx.doi.org/10.1007/BF01217730}{{\em Commun.Math.Phys.} {\bf 121}
  (1989)  351}.

\bibitem{Reshetikhin:1991tc}
N.~Reshetikhin and V.~Turaev, ``{Invariants of three manifolds via link
  polynomials and quantum groups},''
  \href{http://dx.doi.org/10.1007/BF01239527}{{\em Invent.Math.} {\bf 103}
  (1991)  547--597}.

\bibitem{Noui:2004iy}
K.~Noui and A.~Perez, ``{Three dimensional loop quantum gravity: Physical
  scalar product and spin foam models},''
  \href{http://dx.doi.org/10.1088/0264-9381/22/9/017}{{\em Class. Quant. Grav.}
  {\bf 22} (2005)  1739--1762},
\href{http://arxiv.org/abs/gr-qc/0402110}{{\tt arXiv:gr-qc/0402110}}.

\bibitem{Horowitz:1989ng}
G.~T. Horowitz, ``{Exactly Soluble Diffeomorphism Invariant Theories},''
\href{http://dx.doi.org/10.1007/BF01218410}{{\em Commun. Math. Phys.} {\bf 125}
  (1989)  417}.

\bibitem{Crane:1993if}
L.~Crane and D.~Yetter, ``{A Categorical construction of 4-D topological
  quantum field theories},''
\href{http://arxiv.org/abs/hep-th/9301062}{{\tt arXiv:hep-th/9301062}}.

\bibitem{Bonzom:2012mb}
V.~Bonzom and M.~Smerlak, ``{Gauge symmetries in spinfoam gravity: the case for
  'cellular quantization'},''
\href{http://arxiv.org/abs/1201.4996}{{\tt arXiv:1201.4996 [gr-qc]}}.

\bibitem{Barrett:1997gw}
J.~W. Barrett and L.~Crane, ``{Relativistic spin networks and quantum
  gravity},'' \href{http://dx.doi.org/10.1063/1.532254}{{\em J. Math. Phys.}
  {\bf 39} (1998)  3296--3302},
\href{http://arxiv.org/abs/gr-qc/9709028}{{\tt arXiv:gr-qc/9709028}}.

\bibitem{Barrett:1999qw}
J.~W. Barrett and L.~Crane, ``{A Lorentzian signature model for quantum general
  relativity},'' \href{http://dx.doi.org/10.1088/0264-9381/17/16/302}{{\em
  Class. Quant. Grav.} {\bf 17} (2000)  3101--3118},
\href{http://arxiv.org/abs/gr-qc/9904025}{{\tt arXiv:gr-qc/9904025}}.

\bibitem{Engle:2007wy}
J.~Engle, E.~Livine, R.~Pereira, and C.~Rovelli, ``{LQG vertex with finite
  Immirzi parameter},''
  \href{http://dx.doi.org/10.1016/j.nuclphysb.2008.02.018}{{\em Nucl. Phys.}
  {\bf B799} (2008)  136--149},
\href{http://arxiv.org/abs/0711.0146}{{\tt arXiv:0711.0146 [gr-qc]}}.

\bibitem{Freidel:2007py}
L.~Freidel and K.~Krasnov, ``{A New Spin Foam Model for 4d Gravity},''
  \href{http://dx.doi.org/10.1088/0264-9381/25/12/125018}{{\em Class. Quant.
  Grav.} {\bf 25} (2008)  125018},
\href{http://arxiv.org/abs/0708.1595}{{\tt arXiv:0708.1595 [gr-qc]}}.

\bibitem{Livine:2007ya}
E.~R. Livine and S.~Speziale, ``{Consistently Solving the Simplicity
  Constraints for Spinfoam Quantum Gravity},'' {\em Europhys. Lett.} {\bf 81}
  (2008)  50004,
\href{http://arxiv.org/abs/0708.1915}{{\tt arXiv:0708.1915 [gr-qc]}}.

\bibitem{Alexandrov:2010pg}
S.~Alexandrov, ``{The new vertices and canonical quantization},''
  \href{http://dx.doi.org/10.1103/PhysRevD.82.024024}{{\em Phys. Rev.} {\bf
  D82} (2010)  024024},
\href{http://arxiv.org/abs/1004.2260}{{\tt arXiv:1004.2260 [gr-qc]}}.

\bibitem{Alexandrov:2010un}
S.~Alexandrov and P.~Roche, ``{Critical Overview of Loops and Foams},''
  \href{http://dx.doi.org/10.1016/j.physrep.2011.05.002}{{\em Phys. Rept.} {\bf
  506} (2011)  41--86},
\href{http://arxiv.org/abs/1009.4475}{{\tt arXiv:1009.4475 [gr-qc]}}.

\bibitem{Alexandrov:2008da}
S.~Alexandrov, ``{Simplicity and closure constraints in spin foam models of
  gravity},'' \href{http://dx.doi.org/10.1103/PhysRevD.78.044033}{{\em Phys.
  Rev.} {\bf D78} (2008)  044033},
\href{http://arxiv.org/abs/0802.3389}{{\tt arXiv:0802.3389 [gr-qc]}}.

\bibitem{Geiller:2011kr}
M.~Geiller and K.~Noui, ``{Testing the imposition of the Spin Foam Simplicity
  Constraints},''
\href{http://arxiv.org/abs/1112.1965}{{\tt arXiv:1112.1965 [gr-qc]}}.

\bibitem{Alexandrov:2011km}
S.~Alexandrov, M.~Geiller, and K.~Noui, ``{Spin Foams and Canonical
  Quantization},''
\href{http://arxiv.org/abs/1112.1961}{{\tt arXiv:1112.1961 [gr-qc]}}.

\bibitem{Plebanski:1977zz}
J.~F. Plebanski, ``{On the separation of Einsteinian substructures},''
{\em J. Math. Phys.} {\bf 18} (1977)  2511--2520.

\bibitem{DePietri:1998mb}
R.~{DePietri} and L.~{Freidel}, ``{so(4) Plebanski Action and Relativistic Spin
  Foam Model},'' \href{http://dx.doi.org/10.1088/0264-9381/16/7/303}{{\em
  Class. Quant. Grav.} {\bf 16} (1999)  2187--2196},
\href{http://arxiv.org/abs/gr-qc/9804071}{{\tt arXiv:gr-qc/9804071}}.

\bibitem{Dittrich:2008ar}
B.~Dittrich and J.~P. Ryan, ``{Phase space descriptions for simplicial 4d
  geometries},''
\href{http://arxiv.org/abs/0807.2806}{{\tt arXiv:0807.2806 [gr-qc]}}.

\bibitem{Conrady:2008mk}
F.~Conrady and L.~Freidel, ``{On the semiclassical limit of 4d spin foam
  models},'' \href{http://dx.doi.org/10.1103/PhysRevD.78.104023}{{\em Phys.
  Rev.} {\bf D78} (2008)  104023},
\href{http://arxiv.org/abs/0809.2280}{{\tt arXiv:0809.2280 [gr-qc]}}.

\bibitem{Alexandrov:2007pq}
S.~Alexandrov, ``{Spin foam model from canonical quantization},''
  \href{http://dx.doi.org/10.1103/PhysRevD.77.024009}{{\em Phys. Rev.} {\bf
  D77} (2008)  024009},
\href{http://arxiv.org/abs/0705.3892}{{\tt arXiv:0705.3892 [gr-qc]}}.

\bibitem{Engle:2007qf}
J.~Engle, R.~Pereira, and C.~Rovelli, ``{Flipped spinfoam vertex and loop
  gravity},'' \href{http://dx.doi.org/10.1016/j.nuclphysb.2008.02.002}{{\em
  Nucl. Phys.} {\bf B798} (2008)  251--290},
\href{http://arxiv.org/abs/0708.1236}{{\tt arXiv:0708.1236 [gr-qc]}}.

\bibitem{Gielen:2010cu}
S.~Gielen and D.~Oriti, ``{Classical general relativity as BF-Plebanski theory
  with linear constraints},''
  \href{http://dx.doi.org/10.1088/0264-9381/27/18/185017}{{\em Class. Quant.
  Grav.} {\bf 27} (2010)  185017},
\href{http://arxiv.org/abs/1004.5371}{{\tt arXiv:1004.5371 [gr-qc]}}.

\bibitem{Gielen:2011mk}
S.~Gielen and D.~K. Wise, ``{Spontaneously broken Lorentz symmetry for
  Hamiltonian gravity},''
\href{http://arxiv.org/abs/1111.7195}{{\tt arXiv:1111.7195 [gr-qc]}}.

\bibitem{Mondragon:2004nw}
M.~Mondragon and M.~Montesinos, ``{Covariant canonical formalism for
  four-dimensional BF theory},''
  \href{http://dx.doi.org/10.1063/1.2161805}{{\em J.Math.Phys.} {\bf 47} (2006)
   022301}, \href{http://arxiv.org/abs/gr-qc/0402041}{{\tt arXiv:gr-qc/0402041
  [gr-qc]}}.

\bibitem{Alexandrov:2001wt}
S.~Alexandrov, ``{Choice of connection in loop quantum gravity},''
  \href{http://dx.doi.org/10.1103/PhysRevD.65.024011}{{\em Phys. Rev.} {\bf
  D65} (2002)  024011},
\href{http://arxiv.org/abs/gr-qc/0107071}{{\tt arXiv:gr-qc/0107071}}.

\bibitem{Alexandrov:2002br}
S.~Alexandrov and E.~R. Livine, ``{SU(2) loop quantum gravity seen from
  covariant theory},'' \href{http://dx.doi.org/10.1103/PhysRevD.67.044009}{{\em
  Phys. Rev.} {\bf D67} (2003)  044009},
\href{http://arxiv.org/abs/gr-qc/0209105}{{\tt arXiv:gr-qc/0209105}}.

\bibitem{Immirzi:1996dr}
G.~Immirzi, ``{Quantum gravity and Regge calculus},''
  \href{http://dx.doi.org/10.1016/S0920-5632(97)00354-X}{{\em Nucl. Phys. Proc.
  Suppl.} {\bf 57} (1997)  65--72},
\href{http://arxiv.org/abs/gr-qc/9701052}{{\tt arXiv:gr-qc/9701052}}.

\bibitem{Ooguri:1992eb}
H.~Ooguri, ``{Topological lattice models in four-dimensions},''
  \href{http://dx.doi.org/10.1142/S0217732392004171}{{\em Mod. Phys. Lett.}
  {\bf A7} (1992)  2799--2810},
\href{http://arxiv.org/abs/hep-th/9205090}{{\tt arXiv:hep-th/9205090}}.

\bibitem{Ashtekar:1998ak}
A.~Ashtekar, A.~Corichi, and J.~A. Zapata, ``{Quantum theory of geometry. III:
  Non-commutativity of Riemannian structures},''
  \href{http://dx.doi.org/10.1088/0264-9381/15/10/006}{{\em Class. Quant.
  Grav.} {\bf 15} (1998)  2955--2972},
\href{http://arxiv.org/abs/gr-qc/9806041}{{\tt arXiv:gr-qc/9806041}}.

\bibitem{Baratin:2010nn}
A.~Baratin, B.~Dittrich, D.~Oriti, and J.~Tambornino, ``{Non-commutative flux
  representation for loop quantum gravity},''
  \href{http://dx.doi.org/10.1088/0264-9381/28/17/175011}{{\em
  Class.Quant.Grav.} {\bf 28} (2011)  175011},
\href{http://arxiv.org/abs/1004.3450}{{\tt arXiv:1004.3450 [hep-th]}}.

\bibitem{Reisenberger:1997sk}
M.~P. Reisenberger, ``{A Lattice world sheet sum for 4-d Euclidean general
  relativity},''
\href{http://arxiv.org/abs/gr-qc/9711052}{{\tt arXiv:gr-qc/9711052 [gr-qc]}}.

\bibitem{Alexandrov:2000jw}
S.~Alexandrov, ``{SO(4,C)-covariant Ashtekar-Barbero gravity and the Immirzi
  parameter},'' \href{http://dx.doi.org/10.1088/0264-9381/17/20/307}{{\em
  Class. Quant. Grav.} {\bf 17} (2000)  4255--4268},
\href{http://arxiv.org/abs/gr-qc/0005085}{{\tt arXiv:gr-qc/0005085}}.

\bibitem{Baratin:2011hp}
A.~Baratin and D.~Oriti, ``{Group field theory and simplicial gravity path
  integrals: A model for Holst-Plebanski gravity},''
\href{http://arxiv.org/abs/1111.5842}{{\tt arXiv:1111.5842 [hep-th]}}.

\bibitem{Freidel:1999rr}
L.~Freidel, K.~Krasnov, and R.~Puzio, ``{BF description of higher-dimensional
  gravity theories},'' {\em Adv. Theor. Math. Phys.} {\bf 3} (1999)
  1289--1324,
\href{http://arxiv.org/abs/hep-th/9901069}{{\tt arXiv:hep-th/9901069}}.

\bibitem{Barrett:2009gg}
J.~W. Barrett, R.~J. Dowdall, W.~J. Fairbairn, H.~Gomes, and F.~Hellmann,
  ``{Asymptotic analysis of the EPRL four-simplex amplitude},''
  \href{http://dx.doi.org/10.1063/1.3244218}{{\em J. Math. Phys.} {\bf 50}
  (2009)  112504},
\href{http://arxiv.org/abs/0902.1170}{{\tt arXiv:0902.1170 [gr-qc]}}.

\bibitem{Livine:2002ak}
E.~R. Livine, ``{Projected spin networks for Lorentz connection: Linking spin
  foams and loop gravity},''
  \href{http://dx.doi.org/10.1088/0264-9381/19/21/316}{{\em Class. Quant.
  Grav.} {\bf 19} (2002)  5525--5542},
\href{http://arxiv.org/abs/gr-qc/0207084}{{\tt arXiv:gr-qc/0207084}}.

\bibitem{Baratin:2010wi}
A.~Baratin and D.~Oriti, ``{Group field theory with non-commutative metric
  variables},'' \href{http://dx.doi.org/10.1103/PhysRevLett.105.221302}{{\em
  Phys.Rev.Lett.} {\bf 105} (2010)  221302},
\href{http://arxiv.org/abs/1002.4723}{{\tt arXiv:1002.4723 [hep-th]}}.

\bibitem{Rovelli:2010ed}
C.~Rovelli and S.~Speziale, ``{Lorentz covariance of loop quantum gravity},''
  \href{http://dx.doi.org/10.1103/PhysRevD.83.104029}{{\em Phys.Rev.} {\bf D83}
  (2011)  104029}, \href{http://arxiv.org/abs/1012.1739}{{\tt arXiv:1012.1739
  [gr-qc]}}.

\bibitem{Alexandrov:2001pa}
S.~Alexandrov and D.~Vassilevich, ``{Area spectrum in Lorentz covariant loop
  gravity},'' \href{http://dx.doi.org/10.1103/PhysRevD.64.044023}{{\em Phys.
  Rev.} {\bf D64} (2001)  044023},
\href{http://arxiv.org/abs/gr-qc/0103105}{{\tt arXiv:gr-qc/0103105}}.

\bibitem{Buffenoir:2004vx}
E.~Buffenoir, M.~Henneaux, K.~Noui, and P.~Roche, ``{Hamiltonian analysis of
  Plebanski theory},''
  \href{http://dx.doi.org/10.1088/0264-9381/21/22/012}{{\em Class. Quant.
  Grav.} {\bf 21} (2004)  5203--5220},
\href{http://arxiv.org/abs/gr-qc/0404041}{{\tt arXiv:gr-qc/0404041}}.

\bibitem{Alexandrov:2006wt}
S.~Alexandrov, E.~Buffenoir, and P.~Roche, ``{Plebanski theory and covariant
  canonical formulation},''
  \href{http://dx.doi.org/10.1088/0264-9381/24/11/003}{{\em Class. Quant.
  Grav.} {\bf 24} (2007)  2809--2824},
\href{http://arxiv.org/abs/gr-qc/0612071}{{\tt arXiv:gr-qc/0612071}}.

\bibitem{Alexandrov:2008fs}
S.~Alexandrov and K.~Krasnov, ``{Hamiltonian Analysis of non-chiral Plebanski
  Theory and its Generalizations},''
  \href{http://dx.doi.org/10.1088/0264-9381/26/5/055005}{{\em Class. Quant.
  Grav.} {\bf 26} (2009)  055005},
\href{http://arxiv.org/abs/0809.4763}{{\tt arXiv:0809.4763 [gr-qc]}}.

\end{thebibliography}

\end{document}